\documentclass[a4paper,11pt]{article}
\pdfoutput=1 

\usepackage{jheppub} 

\usepackage[T1]{fontenc} 

\usepackage{mathrsfs}
\usepackage{amsmath}
\usepackage{dsfont}
\usepackage{hyperref}

\usepackage{slashed} 
\usepackage{enumitem}

\usepackage{graphicx}
\usepackage{dcolumn}
\usepackage{bm}
\usepackage{upgreek}
\def\p{\textbf{p}}
\def\k{\textbf{k}}

\def\0{\textbf{0}}
\def\x{\textbf{x}}
\def\y{\textbf{y}}

\renewcommand\thefigure{\arabic{figure}}
\renewcommand{\epsilon}{\varepsilon}

\newcommand{\dual}[1]{\overset{\:{}^{^{{{\neg}}}}}{\smash[t]{#1}}} 

\newcommand\varpm{\mathbin{\vcenter{\hbox{%
  \oalign{\hfil$\scriptstyle+$\hfil\cr
          \noalign{\kern-.3ex}
          $\scriptscriptstyle({-})$\cr}%
}}}}
\newcommand\varmp{\mathbin{\vcenter{\hbox{%
  \oalign{$\scriptstyle({+})$\cr
          \noalign{\kern-.3ex}
          \hfil$\scriptscriptstyle-$\hfil\cr}%
}}}}
\allowdisplaybreaks

\title{\boldmath Wigner multiplets in QFT: from Wigner degeneracy to Elko fields}  
\author[a]{Cheng-Yang Lee}
\affiliation[a]{Department of Physics and Chongqing Key Laboratory for Strongly Coupled Physics,
Chongqing University, Chongqing 401331, China} 
\emailAdd{chengyanglee@outlook.com}
\author[b]{Ruifeng Leng}
\affiliation[b]{Department of Physics and Center for Field Theory and Particle Physics, \\Fudan University, Shanghai 200438, China}
\emailAdd{lruifeng@fudan.edu.cn} 
\author[a]{Siyi Zhou}
\emailAdd{siyi@cqu.edu.cn} 

\abstract{
We establish the theoretical foundation of the Wigner superposition field, a quantum field framework for spin-1/2 fermions that exhibit a Wigner doublet --- a discrete quantum number arising from nontrivial representations of the extended Poincar\'{e} group. 
In contrast to the previously developed doublet formalism, which treats the Wigner degeneracy as a superficial label, the superposition formalism encodes it directly into the structure of a unified field via a coherent superposition of degenerate spinor fields. 
By imposing the Lorentz covariance, causality, and canonical quantization, we derive nontrivial constraints on the field configuration, which uniquely identify the Elko field as the consistent realization of the Wigner superposition field. 
Our analysis further clarifies that although the Elko field is a spinor field, it possesses mass dimension one and obeys the Klein-Gordon rather than the Dirac kinematics.
Moreover, we explore the general Elko representation through basis redefinitions, showing that certain traditional properties, such as being eigenspinors of charge conjugation, are artifacts of specific basis choices rather than intrinsic features.
Finally, we discuss the physical implications of Elko as a dark matter (DM) candidate. This work lays the foundation for a systematic reformulation of Elko interactions and its phenomenology as a viable component of DM.

}

\begin{document} 
\maketitle
\flushbottom

\section{Introduction}



\textit{Wigner multiplets in QFT} is a research program whose objective is to develop the theory and explore the phenomenology of nontrivial representations of the extended Poincar\'{e} group. A typical state of the Wigner multiplet is denoted as $|\p,\sigma,n\rangle$ where $\p$ is the momentum, $\sigma$ the spin projection, and $n$ a discrete degree of freedom known as the \textit{Wigner degeneracy}. 
While continuous Lorentz transformations are assumed to have no effects on $n$, discrete transformations can map $|\p,\sigma,n\rangle$ to a superposition state summed over $n$. Since no Standard Model (SM) particle furnishes such a representation, if the Wigner multiplet exists, it must describe an elementary particle beyond the SM.

The study of Wigner multiplets started in Ref.~\cite{Ahluwalia:2022yvk,Ahluwalia:2023slc}, where it is shown that the massive spin-1/2 fermions with the two-fold Wigner degeneracy ($n=\pm\frac{1}{2}$) can be described by Elko, a fermionic field of mass dimension one that satisfies the Klein-Gordon equation but not the Dirac equation~\cite{Ahluwalia:2004ab,Ahluwalia:2004sz,ahluwalia_2019,Ahluwalia:2022ttu}~\footnote{Elko is the German acronym for Eigenspinoren des Ladungskonjugationsoperators. In English, it means eigenspinors of the charge-conjugation operator. The original formulation of the theory does not include the Wigner degeneracy, resulting in a lack of rotational covariance --- a problem later resolved in Ref.~\cite{Ahluwalia:2022yvk,Ahluwalia:2023slc}.}. 
However, it soon became clear that the kinematics of such a fermion is not unique. 
For definitiveness, let $\psi_{+\frac{1}{2}}(x)$ and $\psi_{-\frac{1}{2}}(x)$ be the spinor fields in the $\left(\frac{1}{2},0\right)\oplus\left(0,\frac{1}{2}\right)$ representation corresponding to the $n=+\frac{1}{2}$ and $n=-\frac{1}{2}$ sectors, respectively. 
As we have proposed in Ref.~\cite{Lee:2025kpz}, there are two possible field frameworks for the Wigner doublet:
\begin{enumerate}
    \item Doublet construction: $\Psi(x)\equiv\left[\begin{matrix}
        \psi_{+\frac{1}{2}}(x) \\
        \psi_{-\frac{1}{2}}(x)
    \end{matrix}\right]$.
    \item Superposition over the Wigner degeneracy: $\lambda(x)\equiv \dfrac{1}{\sqrt{2}} \left[ \psi_{+\frac{1}{2}}(x) + \psi_{-\frac{1}{2}}(x) \right]$.
\end{enumerate}
The first paper in this program~\cite{Lee:2025kpz} has established the theoretical foundation of the doublet field formalism. 
In this work, we shall do the same for the superposition formalism. Although Ref.~\cite{Ahluwalia:2022yvk,Ahluwalia:2023slc} have already established locality and Lorentz covariance for the Wigner superposition field in the form of Elko, their analyses did not explore all possible configurations. More precisely, the primary motivation of those works was to show that Elko, or equivalently, the complete set of Majorana spinors, can give rise to a consistent QFT.

Elko originated from the desire to understand the physics of Majorana spinors and neutrinos~\cite{Ahluwalia:1993cz,Ahluwalia:1993pf,Ahluwalia:2003jt}. It was therefore a surprise to find that Majorana spinors bear no relation to neutrinos. When treated as spinors in momentum space and not as anti-commuting Grassmann variables, the Majorana spinors do not satisfy the Dirac equation. Instead, the QFT constructed using Majorana spinors was found to have mass dimension one and Klein-Gordon kinematics. Due to its mass dimension, the fermion has renormalizable self-interactions, making it a viable candidate for self-interacting DM. Since the original works of Ahluwalia and Grumiller~\cite{Ahluwalia:2004ab,Ahluwalia:2004sz}, Elko has been extensively studied in mathematical physics~\cite{daRocha:2005ti,daRocha:2008we,HoffdaSilva:2012uke,daRocha:2013qhu,Bonora:2014dfa,Cavalcanti:2014wia,daRocha:2016bil,HoffdaSilva:2017waf,Fabbri:2017lvu,Meert:2018qzk,Arcodia:2019flm,Rogerio:2020trs,Rogerio:2020cqg,Rogerio:2021ewp,Rogerio:2022tsl}, cosmology~\cite{Basak:2014qea,Boehmer:2006qq,Boehmer:2007dh,Boehmer:2008ah,Boehmer:2008rz,Boehmer:2009aw,Boehmer:2010ma,HoffdaSilva:2014tth,Pereira:2014wta,S:2014woy,Pereira:2014pqa,Pereira:2016emd,Pereira:2016eez,Pereira:2017efk,Pereira:2017bvq,BuenoRogerio:2017zxf,Pereira:2018xyl,Pereira:2018hir,Pereira:2020ogo,Pereira:2021dkn,Lima:2022vrc}, brane worlds~\cite{Jardim:2014xla,Dantas:2015mfi,Zhou:2017bbj,ZhouZhouXiangNan:2018het,Sorkhi:2018jhy,MoazzenSorkhi:2020fqp} and DM~\cite{Dias:2010aa,Agarwal:2014oaa,Alves:2014kta,Alves:2014qua,Alves:2017joy,Moura:2021rmf}. However, these works were completed prior to the introduction of Wigner degeneracy, which resolved the long-standing problem of rotation symmetry. As a result, many results and interpretations have to be re-examined. The task we are undertaking here is the first step in this direction. What we will accomplish here is to establish the foundation of the Wigner superposition field. 
Starting with the most generic spinor fields in the superposition formalism, we impose constraints from discrete symmetries, canonical quantization, and the physical Hamiltonian. In doing so, we uncover a crucial distinguishing feature of the superposition formalism from the doublet construction --- \textit{the relative phases between spinors}. As a result, we demonstrate that Elko emerges as the unique consistent realization within the superposition framework.

This paper is organized as follows. In Section~\ref{wigner_spinor_field}, we present the most generic spinor fields with the two-fold Wigner degeneracy that respect Lorentz symmetry in the $\left(\frac{1}{2},0\right)\oplus\left(0,\frac{1}{2}\right)$ representation. The degenerate degrees of freedom $n=\pm\frac{1}{2}$ and the demand of locality yield a nontrivial relation for the spinor phases. That is, the locality structure of the fields bifurcates into the scenarios of \textit{causality non-mixing} and \textit{mixing}. In Section~\ref{sec:Dirac_0} and Section~\ref{sec:elko_0}, we compute the canonical anti-commutators of $\lambda(x)$ with its Dirac dual $\bar{\lambda}(x)$ and Elko dual $\dual{\lambda}(x)$ fields. In Section~\ref{sec:basisT_0}, we show that the \textit{causality non-mixing} and \textit{mixing} scenarios are related by a basis transformation. 
In Section~\ref{sec:pct}, we analyze the discrete transformations of the superposition fields and provide the condition for the existence of a physical Wigner doublet. 
In Section~\ref{sec:can_W_s_1}, we perform the canonical quantization of the superposition fields. Requiring that the fields satisfy the canonical anti-commutation relations and yield a positive-definite (free) Hamiltonian uniquely singles out the Elko field as the consistent solution. 
Finally, we present the basis transformation that connects the spinors constructed in this work to those commonly used in the literature and demonstrate their physical consistency.


\section{A general spinor field with the Wigner degeneracy}\label{wigner_spinor_field}

In this section, we establish a fundamental framework for quantum spinor fields describing a massive spin-1/2 fermion doublet, which has internal degrees of freedom, including the usual spin projection $\sigma=\pm\frac{1}{2}$ and, in particular, the two-fold Wigner degeneracy $w=\frac{1}{2}, \ n=\pm\frac{1}{2}$. 
Thus, a Wigner-degenerate fermion must be an excitation of a quantum spinor field that incorporates both the spin projection and the Wigner degeneracy as fundamental components of its degrees of freedom.
Since a detailed description has been presented in our previous work~\cite{Lee:2025kpz}, we will just provide crucial results and expressions here.

\subsection{The two-fold Wigner-degenerate fields}
\label{sec:Spin_m_1}

The two spinor fields $\psi_{n}(x)$, with mass dimension $[\psi_{n}]=\frac{3}{2}$, correspond to the two Wigner degeneracies $n=\pm \frac{1}{2}$, and can be constructed in a general form based on the requirement of Poincar\'{e} covariance 
\begin{align}
\psi_{n, \ell}(x) &= \psi_{n, \ell}^{+}(x) + \psi_{n, \ell}^{-c}(x)
\, ,
\label{eq:sf_W_20}
\end{align}
with
\begin{align}
\psi_{n, \ell}^{+}(x)
&=
\int\dfrac{d^{3}p}{(2\pi)^{3}}\frac{1}{\sqrt{2E_{\p}}}\sum_{\sigma} e^{-ip\cdot x}u_{n, \ell}(\p,\sigma)a_{n}(\p,\sigma) 
\, , 
\label{eq:sf_W_20+} \\
\psi_{n, \ell}^{-c}(x)
&=
\int\dfrac{d^{3}p}{(2\pi)^{3}}\frac{1}{\sqrt{2E_{\p}}}\sum_{\sigma} e^{ip\cdot x} v_{n, \ell}(\p,\sigma)a^{c \dag}_{n}(\p,\sigma) 
\label{eq:sf_W_20-} \, ,
\end{align}
where polarizations $u_{n}(\p,\sigma)$, $v_{n}(\p,\sigma)$ are four-component Dirac spinors.
$a_{n}^{(\dag)}(\p,\sigma)$ and $a^{c(\dag)}_{n}(\p,\sigma)$ denote the annihilation (creation) operators for the particle and its associated antiparticle respectively, both having the same mass $m$. The two-fold physical states for the Wigner-degenerate particle and its antiparticle are defined as~\footnote{We use the bracket to denote an alternative item. For instance, in Eq.~\eqref{eq:deg_op_ab}, the expression can be interpreted using either $a, a_{n}^{\dag}$ or $a^{c}, a_{n}^{c \dag}$.}
\begin{align}
|\p,\sigma, n; a^{(c)} \rangle \equiv \sqrt{2 E_{\p}} \,  a_{n}^{(c) \dag}(\p, \sigma) |0\rangle \, ,
 \quad 
\text{with} \quad 
n=\pm\frac{1}{2} \, ,
\label{eq:deg_op_ab}
\end{align}
where $\p$ refers to the physical momentum, and $\sigma$ runs over the two independent spin projections $\pm\frac{1}{2}$.
These on-shell momentum eigenstates are Lorentz-invariantly ortho-normalized as
\begin{equation}
    \langle \p^{\prime},\sigma^{\prime}, n'; a^{(c)} |\p,\sigma, n; a^{(c)} \rangle 
    = 2 E_{\p} \left(2\pi \right)^{3} \delta^{(3)} \left( \p^{\prime} - \p \right) \delta_{\sigma^{\prime} \sigma} \delta_{n' n} \, .
\label{eq:norm_1p_W_2}
\end{equation}
The Wigner-degenerate states transform under the Lorentz group with the Wigner degeneracy index $n$ remaining invariant.
The fermionic creation and annihilation operators for the Wigner doublet satisfy the canonical quantization relations
\begin{align}
\bigl\{a_{n}(\p, \sigma), a_{n'}^{\dag}(\p^{\prime},\sigma')\bigr\}
&=\bigl\{a^{c}_{n}(\p,\sigma), a^{c \dag}_{n'}(\p^{\prime},\sigma') \bigr\}
=(2 \pi)^{3} \delta^{(3)}(\p-\p') \delta_{\sigma\sigma'} \delta_{nn'}
\, ,
\label{eq:can_W_ab_1} \\
\bigl\{a_{n}(\p, \sigma), a_{n'}^{c}(\p^{\prime},\sigma')\bigr\}
&= \bigl\{a_{n}^{\dag}(\p, \sigma), a_{n'}^{c \dag}(\p^{\prime},\sigma')\bigr\}
= \bigl\{a_{n}(\p,\sigma), a^{c \dag}_{n'}(\p^{\prime},\sigma') \bigr\}
=0
\, .
\label{eq:can_W_ab_2}
\end{align}
The free vacuum $|0\rangle$ is defined to be the state in Hilbert space such that
\begin{align}
 a_{n}(\p, \sigma) |0\rangle = a_{n}^{c}(\p, \sigma) |0\rangle = 0 \, .
\label{eq:vac_W_ab_1}
\end{align}
The general solutions for the polarizations $u_{n}(\p,\sigma)$, $v_{n}(\p,\sigma)$~\eqref{eq:sf_W_20} can be determined by 
the requirement that both Lorentz transformations and discrete inversions map the Wigner-degenerate fields at a given spacetime point into superpositions of themselves at the corresponding point. 
They are eigenstates of $p_{\mu} \gamma^{\mu}$ and thus solutions of the momentum space Dirac equation up to an additional factor
\begin{align}
p_{\mu} \gamma^{\mu} u_{n}(\p,\sigma) &= m b_{u, n} \, u_{n}(\p,\sigma) \, ,
\label{eq:Dirac_eq_u_gen_1} \\
p_{\mu} \gamma^{\mu} v_{n}(\p,\sigma) &= m b_{v, n} \, v_{n}(\p,\sigma)
\, ,
\label{eq:Dirac_eq_v_gen_1}
\end{align}
where $b_{u, n}, b_{v, n} = \pm 1$ are sign factors. Taking the Dirac adjoint yields 
\begin{align}
\bar{u}_{n}(\p,\sigma) p_{\mu} \gamma^{\mu} &= m b_{u, n} \, \bar{u}_{n}(\p,\sigma) \, ,
\label{eq:Dirac_eq_u_gen_2} \\
\bar{v}_{n}(\p,\sigma) p_{\mu} \gamma^{\mu} &= m b_{v, n} \, \bar{v}_{n}(\p,\sigma)
\, ,
\label{eq:Dirac_eq_v_gen_2}
\end{align}
with the Dirac dual spinors defined as $\bar{u}_{n} \equiv u^{\dag}_{n} \gamma^{0}$ and $\bar{v}_{n} \equiv v^{\dag}_{n} \gamma^{0}$.
In the rest frame, after adjusting the overall mass scale according to $\psi_{n}$~\eqref{eq:sf_W_20} and ensuring consistency with the normalization of states in Eq.~\eqref{eq:norm_1p_W_2}, polarizations can be expressed explicitly as
\begin{alignat}{2}
u_{+\frac{1}{2}}(\0,\textstyle{+\frac{1}{2}})&=\sqrt{m}\left[\begin{matrix}
    1 \\
    0 \\
    b_{u,+\frac{1}{2}} \\
    0
\end{matrix}\right] \, ,\quad
&u_{+\frac{1}{2}}(\0,\textstyle{-\frac{1}{2}})&=\sqrt{m}\left[\begin{matrix}
    0 \\
    1 \\
    0 \\
    b_{u,+\frac{1}{2}}
\end{matrix}\right] \, ,\label{eq:+u}\\
v_{+\frac{1}{2}}(\0,\textstyle{+\frac{1}{2}})&=\sqrt{m}\left[\begin{matrix}
    0 \\
    1 \\
    0 \\
    b_{v,+\frac{1}{2}}
\end{matrix}\right] \, ,\quad
&v_{+\frac{1}{2}}(\0,\textstyle{-\frac{1}{2}})&=\sqrt{m}\left[\begin{matrix}
    -1 \\
    0 \\
    -b_{v,+\frac{1}{2}} \\
    0
\end{matrix}\right] \, .\label{eq:+v}\\
u_{-\frac{1}{2}}(\0,\textstyle{+\frac{1}{2}})&=\sqrt{m}\left[\begin{matrix}
    1 \\
    0 \\
    b_{u,-\frac{1}{2}} \\
    0
\end{matrix}\right] \, ,\quad
&u_{-\frac{1}{2}}(\0,\textstyle{-\frac{1}{2}})&=\sqrt{m}\left[\begin{matrix}
    0 \\
    1 \\
    0 \\
    b_{u,-\frac{1}{2}}
\end{matrix}\right] \, ,\label{eq:-u}\\
v_{-\frac{1}{2}}(\0,\textstyle{+\frac{1}{2}})&=\sqrt{m}\left[\begin{matrix}
    0 \\
    1 \\
    0 \\
    b_{v,-\frac{1}{2}}
\end{matrix}\right] \, ,\quad
&v_{-\frac{1}{2}}(\0,\textstyle{-\frac{1}{2}})&=\sqrt{m}\left[\begin{matrix}
    -1 \\
    0 \\
    -b_{v,-\frac{1}{2}} \\
    0
\end{matrix}\right] \, . \label{eq:-v}
\end{alignat}
Then, the polarization spinors at an arbitrary momentum $\p$ are obtained through a Lorentz boost
\begin{align}
u_{n}(\p,\sigma) = \mathscr{D}(L(\p)) u_{n}(\0,\sigma) \, , 
\quad
v_{n}(\p,\sigma) = \mathscr{D}(L(\p)) v_{n}(\0,\sigma) \, . 
\label{eq:uv_p_0} 
\end{align}
where $\mathscr{D}(L(\p))$ is the spinor representation (non-unitary) of some standard Lorentz boost $L(\p)$, which is defined by $p^{\mu}={L^{\mu}}_{\nu}(\p)k^{\nu}$ with $k^{\mu} = \left(m, 0, 0, 0 \right)$.
We can then obtain the one-particle state of momentum $\p$ as
\begin{align}
|\p,\sigma, n; a^{(c)} \rangle &= U(L(\p))|\k,\sigma, n; a^{(c)} \rangle \, ,
\label{Lorentz_part_stand_1}
\end{align}
where the unitary operator $U(L(\p))$ is the standard Lorentz boost acting on the Hilbert space.
In the chiral representation, $\mathscr{D}(L(\p))$ is given in an explicit formula
\begin{align}
\mathscr{D}(L(\p)) &=
\sqrt{\dfrac{E_{\p} + m}{2 m}}
\left[\begin{matrix}
    \mathds{1}-\frac{\boldsymbol{\sigma} \cdot \p}{E_{\p} + m} & 0 \\
   0 & \mathds{1}+\frac{\boldsymbol{\sigma} \cdot \p}{E_{\p} + m}
\end{matrix}\right]
\, ,
\label{mat:L_p_spinor_0} 
\end{align}
and satisfies the following properties
\begin{align}
L^{\mu}_{\ \nu}(\p) \gamma^{\nu} &=
\mathscr{D}^{-1}(L(\p)) \gamma^{\mu} \mathscr{D}(L(\p)) \, ,
\label{eq:L_p_spinor_0} \\
\frac{p_{\mu} \gamma^{\mu}}{m} &=
\mathscr{D}(L(\p)) \gamma^{0} \mathscr{D}^{-1}(L(\p)) \, ,
\label{eq:L_p_spinor_00} \\
\mathscr{D}^{\dag}(L(\p)) &=
\gamma^{0} \mathscr{D}^{-1}(L(\p)) \gamma^{0} \, ,
\label{eq:L_p_spinor_01} \\
\mathscr{D}(L(-\p)) &=
\gamma^{0} \mathscr{D}(L(\p)) \gamma^{0} \, ,
\label{eq:L_p_spinor_02} \\
\mathscr{D}^{*}(L(-\p)) &=
\gamma^{5} \mathcal{C} \mathscr{D}(L(\p)) \mathcal{C}^{-1} \gamma^{5} 
= \gamma^{3} \gamma^{1} \mathscr{D}(L(\p)) \gamma^{1} \gamma^{3}
\, ,
\label{eq:L_p_spinor_03} \\
\mathscr{D}^{*}(L(\p)) &=
\gamma^{0} \mathcal{C} \mathscr{D}(L(\p)) \mathcal{C}^{-1} \gamma^{0} 
= -\gamma^{2} \mathscr{D}(L(\p)) \gamma^{2}
\, ,
\label{eq:L_p_spinor_04}
\end{align}
with the $4 \times 4$ Dirac matrix $\mathcal{C} \equiv -i \gamma^{2} \gamma^{0} = -\mathcal{C}^{-1} = -\mathcal{C}^{\dag} = -\mathcal{C}^{\text{T}}$.
Putting together all the above explicit expressions of general polarizations, we find the Lorentz invariant ortho-normalization relations with the Dirac dual spinors  
\begin{align}
\bar{u}_{n}(\p,\sigma) u_{n'}(\p,\sigma') &= m \delta_{\sigma\sigma'} (b_{u, n} + b_{u, n'}) \, ,
\label{eq:norm_Dirac_u_1} \\
\bar{v}_{n}(\p,\sigma) v_{n'}(\p,\sigma') &= m \delta_{\sigma\sigma'} (b_{v, n} + b_{v, n'}) \, ,
\label{eq:norm_Dirac_v_1} \\
\bar{u}_{n}(\p,\sigma) v_{n'}(\p,\sigma) &= \bar{v}_{n}(\p,\sigma) u_{n'}(\p,\sigma) 
\nonumber \\
&= u^{\dag}_{n}(\p,\sigma) v_{n'}(-\p,\sigma) 
= v^{\dag}_{n}(\p,\sigma) u_{n'}(-\p,\sigma) = 0 \, ,
\label{eq:norm_Dirac_uv_m_1} \\
\bar{u}_{n}(\p,\sigma) v_{n'}(\p,-\sigma)  
&= \bar{v}_{n'}(\p,-\sigma) u_{n}(\p,\sigma)
= (-1)^{\frac{1}{2}+\sigma} m (b_{u, n} + b_{v, n'}) \, ,
\label{eq:norm_Dirac_uv_m_2} \\
u^{\dag}_{n}(\p,\sigma) v_{n'}(-\p,-\sigma) 
&= v^{\dag}_{n'}(-\p,-\sigma) u_{n}(\p,\sigma)
= (-1)^{\frac{1}{2}+\sigma} m (1+b_{u, n} b_{v, n'}) \, ,
\label{eq:norm_Dirac_uv_m_3}
\end{align}
as well as the following identities:
\begin{alignat}{2}
\gamma^{0} u_{n}(\p,\sigma)&=b_{u, n} u_{n}(-\p,\sigma) \, , \quad 
&\gamma^{0} v_{n}(\p,\sigma)&= b_{v, n} v_{n}(-\p,\sigma) \, , \label{eq:id_Dirac_uv_01} \\
\gamma^{3} \gamma^{1} u_{n}(\p,\sigma)&=(-1)^{\frac{1}{2}-\sigma} u^{*}_{n}(-\p,-\sigma) \, , \ 
&\gamma^{3} \gamma^{1} v_{n}(\p,\sigma)&=(-1)^{\frac{1}{2}-\sigma} v^{*}_{n}(-\p,-\sigma) \, , \label{eq:id_Dirac_uv_02} \\
u_{n}(\p,\sigma)&= \left( P_{L} + \Delta_{u} P_{R} \right) u_{-n}(\p,\sigma) \, , \ 
&v_{n}(\p,\sigma)&= \left( P_{L} + \Delta_{v} P_{R} \right) v_{-n}(\p,\sigma) \, , \label{eq:id_Dirac_uv_03} 
\end{alignat}
\begin{align}
-i\gamma^{2}u_{n}^{*}(\p,\sigma) &= \left( b_{u,n} P_{L} - b_{v,\pm n} P_{R} \right) v_{\pm n}(\p,\sigma) \, , 
\label{eq:id_Dirac_u_04} \\
-i\gamma^{2}v_{n}^{*}(\p,\sigma) &= \left( b_{u,\pm n} P_{R} - b_{v,n} P_{L} \right) u_{\pm n}(\p,\sigma) \, , \label{eq:id_Dirac_v_04}
\end{align}
where the products of sign factors in Eqs.~\eqref{eq:+u}-\eqref{eq:-v} are denoted as
\begin{align}
\Delta_{u(v)} &\equiv \prod_{n} b_{u(v),n} \, .
\label{eq:prod_b_uv_0} 
\end{align}
We also have the sum of spin projections
\begin{align}
\sum_{\sigma} u_{n} (\p,\sigma) \bar{u}_{n} (\p,\sigma) &= \slashed{p} + b_{u, n} m \mathds{1} \, ,
\label{eq:sum_Dirac_u_0} 
\\
\sum_{\sigma} v_{n} (\p,\sigma) \bar{v}_{n} (\p,\sigma) &= \slashed{p} + b_{v, n} m \mathds{1} \, .
\label{eq:sum_Dirac_v_0}
\end{align}

\subsection{Dirac dual \& Dirac correlators}
\label{sec:Dirac_0}

As highlighted in our previous work~\cite{Lee:2025kpz}, there are two reasonable approaches to fully capturing the physics of two-fold Wigner degeneracy: the doublet construction and the superposition over the Wigner degeneracy. Having thoroughly investigated the doublet framework, in this paper, we focus on the second approach by constructing a two-fold Wigner superposition spinor field:
\begin{align}
\lambda (x)
&\equiv \dfrac{1}{\sqrt{2}} \left[ \psi_{+\frac{1}{2}}(x) + \psi_{-\frac{1}{2}}(x) \right]
\nonumber \\
&=\int\dfrac{d^{3}p}{(2\pi)^{3}}\frac{1}{2\sqrt{E_{\p}}}\sum_{n,\sigma}\left[e^{-ip\cdot x}u_{n}(\p,\sigma)a_{n}(\p,\sigma)+e^{ip\cdot x} v_{n}(\p,\sigma)a^{c \dag}_{n}(\p,\sigma)\right] 
\, ,
\label{eq:field_1+1_W_0}
\end{align}
where $n$ runs over the Wigner degeneracy, and a normalization factor $\sqrt{2}$ has been inserted for future convenience.
The Dirac dual field with respect to $\lambda (x)$ is locally defined as
\begin{align}
\bar{\lambda} (x) \equiv \lambda^{\dag} (x) \gamma^{0}
\, .  
\label{eq:Dirac_dual_0}
\end{align}
From the Fourier expansion of the Wigner superposition field $\lambda (x)$~\eqref{eq:field_1+1_W_0}, the Dirac dual field $\bar{\lambda} (x)$ can be quantized by
\begin{align}
\bar{\lambda} (x) 
&= \int\dfrac{d^{3}p}{(2\pi)^{3}}\frac{1}{2\sqrt{E_{\p}}}\sum_{n,\sigma} \left[ e^{ip\cdot x} \bar{u}_{n}(\p,\sigma)a_{n}^{\dag}(\p,\sigma)
    + e^{-ip\cdot x} \bar{v}_{n}(\p,\sigma)a^{c}_{n}(\p,\sigma) \right]  
\label{eq:Dirac_dual_1}
\, .
\end{align}
Causality is embedded in the two-point correlation function of the local fields $\lambda (x)$ and $\bar{\lambda} (x)$, given by
\begin{align}
\left\{ \lambda (x), \bar{\lambda} (y) \right\}
&= i \gamma^{\mu} \partial^{x}_{\mu} \left[ D(x-y) - D(y-x) \right] 
\nonumber \\
&\quad \ + \dfrac{m}{2} \left[\left(b_{u,+\frac{1}{2}} + b_{u,-\frac{1}{2}} \right) D(x-y) 
+ \left(b_{v,+\frac{1}{2}} + b_{v,-\frac{1}{2}} \right) D(y-x) \right]
\, ,  
\label{eq:casual_1}
\end{align}
where 
\begin{align}
D(x-y) = \int \dfrac{d^{3}p}{(2\pi)^{3}}\frac{1}{2E_{\p}} e^{-ip \cdot (x-y)}
\, ,  
\end{align}
is even for a space-like interval $x-y$. 
To ensure causality, the anticommutator~\eqref{eq:casual_1} must vanish for spacelike separations, which requires that
\begin{align}
b_{u} + b_{v} = 0 
\, ,
\quad \text{with} \quad
 b_{u(v)} \equiv \sum_{n} b_{u(v),n} \, .
\label{eq:causal_Dirac_b_0}
\end{align}
We also notice that the causality condition~\eqref{eq:causal_Dirac_b_0} induces the identity via the sign products~\eqref{eq:prod_b_uv_0}
\begin{align}
\Delta \equiv \Delta_{u} = \Delta_{v} = \pm 1 \, .
\label{eq:causal_Dirac_b_00}
\end{align}
Then, taking the sum over the Wigner degeneracy in Eqs.~\eqref{eq:sum_Dirac_u_0}-\eqref{eq:sum_Dirac_v_0} yields
\begin{align}
\sum_{n,\sigma} u_{n} (\p,\sigma) \bar{u}_{n} (\p,\sigma) &= 2\slashed{p} + b_{u} m \mathds{1} \, ,
\label{eq:sum_Dirac_u_1} \\
\sum_{n,\sigma} v_{n} (\p,\sigma) \bar{v}_{n} (\p,\sigma) &= 2\slashed{p} - b_{u} m \mathds{1} \, .
\label{eq:sum_Dirac_v_1}
\end{align}

If each Wigner-degenerate field $\psi_{n}(x)$ individually preserves the causality condition $\{\psi_{n}(x),\bar{\psi}_{n}(y)\}=0$ for space-like interval $x-y$, the general causality condition of Eq.~\eqref{eq:causal_Dirac_b_0} can be further constrained as
\begin{align}
b_{u,+\frac{1}{2}} + b_{v,+\frac{1}{2}} = b_{u,-\frac{1}{2}} + b_{v,-\frac{1}{2}} = 0 
\, .
\label{eq:causal_Dirac_b_1}
\end{align}
In contrast, if the causality condition can only be preserved by the superposition field $\lambda (x)$~\eqref{eq:field_1+1_W_0} due to a non-zero causality mixing term in Eq.~\eqref{eq:casual_1},
we must instead impose   
\begin{align}
b_{u, +\frac{1}{2}} &= b_{v, +\frac{1}{2}} = -b_{u, -\frac{1}{2}} = -b_{v, -\frac{1}{2}} = \pm1 
\, .
\label{eq:causal_Dirac_b_2}
\end{align}
An important result to note is that the Elko condition, defined as
\begin{align}
b_{u} = b_{v} = 0 \, , 
\label{eq:causal_elko_b_000}
\end{align}
implying
\begin{align}
\Delta = -1 \, ,
\label{eq:causal_elko_b_001}
\end{align}
is allowed in both causality scenarios --- whether or not the causality mixing is present, as outlined in Eqs.~\eqref{eq:causal_Dirac_b_1} and \eqref{eq:causal_Dirac_b_2} respectively. 
Under this condition, the polarization sums in Eqs.~\eqref{eq:sum_Dirac_u_1}-\eqref{eq:sum_Dirac_v_1} reduce to
\begin{align}
\sum_{n,\sigma} u_{n} (\p,\sigma) \bar{u}_{n} (\p,\sigma) &= 2\slashed{p} \, ,
\label{eq:sum_Dirac_u_23} \\
\sum_{n,\sigma} v_{n} (\p,\sigma) \bar{v}_{n} (\p,\sigma) &= 2\slashed{p} \, .
\label{eq:sum_Dirac_v_23}
\end{align}
The Wigner superposition field $\lambda (x)$~\eqref{eq:field_1+1_W_0}, which satisfies the Elko condition, is referred to as the Elko field. A more detailed discussion of its properties and implications within the Klein-Gordon framework will be presented in Section~\ref{sec:elko_KG_1}.

\subsection{Elko dual \& Elko correlators}
\label{sec:elko_0}

As an alternative to the Dirac dual field, which is defined by local fields without derivatives in Eq.~\eqref{eq:Dirac_dual_1},
the Elko dual field for the spinor field $\lambda (x)$~\eqref{eq:field_1+1_W_0} is defined by its total derivatives as~\cite{deGracia:2024umr}
\begin{align}
\dual{\lambda} (x) 
&\equiv \left[ i m^{-1} \gamma^{\mu} \partial_{\mu} \lambda (x) \right]^{\dagger} \gamma^{0}
= -i m^{-1} \partial_{\mu} \bar{\lambda} (x) \gamma^{\mu} 
\, ,
\label{eq:elko_dual_0}
\end{align}
where the second equality shows its relationship with the Dirac dual field. One may be puzzled as to why we have introduced a new dual field. The reason for introducing $\dual{\lambda}(x)$ is to canonically quantize Elko. Here, we present the basic properties of the Elko dual. In Section~\ref{sec:can_W_s_1}, we show that the kinematics of the Elko field is uniquely described by a Klein-Gordon Lagrangian with $\lambda(x)$ and $\dual{\lambda}(x)$ being the canonical variables.

The Elko dual field $\dual{\lambda} (x)$ is Lorentz covariant. Using the Fourier expansion of the Wigner superposition field $\lambda (x)$~\eqref{eq:field_1+1_W_0}, the Elko dual field $\dual{\lambda} (x)$~\eqref{eq:elko_dual_0} can be quantized by the annihilation and creation operators~\eqref{eq:can_W_ab_1}-\eqref{eq:can_W_ab_2} as 
\begin{align}
\dual{\lambda} (x) 
&= \int\dfrac{d^{3}p}{(2\pi)^{3}}\frac{1}{2\sqrt{E_{\p}}}\sum_{n,\sigma} \left[ e^{ip\cdot x} \dual{u}_{n}(\p,\sigma)a_{n}^{\dag}(\p,\sigma)
    + e^{-ip\cdot x} \dual{v}_{n}(\p,\sigma)a^{c}_{n}(\p,\sigma) \right]  
\label{eq:elko_dual_1}
\, ,
\end{align}
where the associated Elko dual polarizations are defined as 
\begin{align}
\dual{u}_{n} (\p,\sigma) &\equiv \left[m^{-1} p_{\mu} \gamma^{\mu} u_{n} (\p,\sigma) \right]^{\dagger} \gamma^{0}
= b_{u, n} \, \bar{u}_{n} (\p,\sigma)
\, , 
\label{eq:elko_dual_u_1} \\
\dual{v}_{n} (\p,\sigma) &\equiv -\left[m^{-1} p_{\mu} \gamma^{\mu} v_{n} (\p,\sigma) \right]^{\dagger} \gamma^{0}
= -b_{v, n} \, \bar{v}_{n} (\p,\sigma)
\, . 
\label{eq:elko_dual_v_1}
\end{align}
Thus, the Elko dual polarizations satisfy the same equations as the Dirac dual polarizations in Eqs.~\eqref{eq:Dirac_eq_u_gen_2}-\eqref{eq:Dirac_eq_v_gen_2}, ensuring their consistency within the Lorentz covariant framework.
Applying for Eqs.~\eqref{eq:norm_Dirac_u_1}-\eqref{eq:norm_Dirac_uv_m_3}, the Lorentz invariant ortho-normalization relations with the Elko dual spinors are
\begin{align}
\dual{u}_{n}(\p,\sigma) u_{n'}(\p,\sigma') &= m \delta_{\sigma\sigma'} (1+b_{u, n} b_{u, n'}) \, ,
\label{eq:norm_elko_u_1} \\
\dual{v}_{n}(\p,\sigma) v_{n'}(\p,\sigma') &= -m \delta_{\sigma\sigma'} (1+b_{v, n} b_{v, n'}) \, ,
\label{eq:norm_elko_v_1} \\
\dual{u}_{n}(\p,\sigma) v_{n'}(\p,\sigma) &= \dual{v}_{n}(\p,\sigma) u_{n'}(\p,\sigma) 
\nonumber \\
&= u^{\dag}_{n}(\p,\sigma) v_{n'}(-\p,\sigma) 
= v^{\dag}_{n}(\p,\sigma) u_{n'}(-\p,\sigma) = 0 \, ,
\label{eq:norm_elko_uv_m_1} \\
\dual{u}_{n}(\p,\sigma) v_{n'}(\p,-\sigma)  
&= \dual{v}_{n'}(\p,\sigma) u_{n}(\p,-\sigma)
\nonumber \\
&=u^{\dag}_{n}(\p,\sigma) v_{n'}(-\p,-\sigma) 
= v^{\dag}_{n'}(-\p,-\sigma) u_{n}(\p,\sigma)
\nonumber \\
&= (-1)^{\frac{1}{2}+\sigma} m (1+b_{u, n} b_{v, n'}) \, .
\label{eq:norm_elko_uv_m_3}
\end{align}
Moreover, the sum of spin projections can be derived from Eqs.~\eqref{eq:sum_Dirac_u_0}-\eqref{eq:sum_Dirac_v_0} and \eqref{eq:elko_dual_0}
\begin{align}
\sum_{\sigma}
u_{n} (\p,\sigma) \dual{u}_{n} (\p,\sigma)
&= b_{u, n} \slashed{p} + m \mathds{1}
\, ,
\label{eq:sum_elko_u_1} \\
\sum_{\sigma}
v_{n} (\p,\sigma) \dual{v}_{n} (\p,\sigma)
&= -b_{v, n} \slashed{p} - m \mathds{1}
\, .
\label{eq:sum_elko_v_1}
\end{align}
As an alternative to using the Dirac dual field in Eq.~\eqref{eq:casual_1}, one can construct the two-point correlation function with the Elko dual field $\dual{\lambda} (x)$~\eqref{eq:elko_dual_1} 
\begin{align}
\left\{ \lambda_{\ell} (x), \dual{\lambda}_{\ell'} (y) \right\}
&= m \left[ D(x-y) - D(y-x) \right] \delta_{\ell \ell'}
\nonumber \\
&\quad \  + \dfrac{i}{2} \, \partial_{\mu}^{x} \sum_{n}
\left[ b_{u, n} \, D(x-y) + b_{v, n} \, D(y-x) \right]
\gamma^{\mu}_{\ell \ell'} 
\, ,  
\label{eq:casual_1+1_elko_0}
\end{align}
with~\footnote{Notice that $\bigl\{ \psi_{n, \ell} (x), \dual{\psi}_{n', \ell'} (y) \bigr\} = 0$ for $n \neq n'$ due to anti-commuting relations for different Wigner degeneracies~\eqref{eq:can_W_ab_1}-\eqref{eq:can_W_ab_2}.}
\begin{align}
\left\{ \psi_{n, \ell} (x), \dual{\psi}_{n, \ell'} (y) \right\}
&= i \partial_{\mu}^{x} \left[ b_{u, n} \, D(x-y) + b_{v, n} \, D(y-x) \right] \gamma^{\mu}_{\ell \bar{\ell}}
\nonumber \\
&\quad + m \left[ D(x-y) - D(y-x) \right] \delta_{\ell \bar{\ell}}
\, ,  
\end{align}
where we have inserted the sum of spin projections in Eqs.~\eqref{eq:sum_elko_u_1}-\eqref{eq:sum_elko_v_1}.
Thus, the causality condition induced by the Elko correlation function of Eq.~\eqref{eq:casual_1+1_elko_0} gives
\begin{align}
b_{u} + b_{v}  = 0 
\, ,
\label{eq:causal_elko_b_0}
\end{align}
which is identical to that derived by the Dirac correlators~\eqref{eq:casual_1} in Eq.~\eqref{eq:causal_Dirac_b_0}.
Imposing this causality condition, the sum of Eqs.~\eqref{eq:sum_elko_u_1}-\eqref{eq:sum_elko_v_1} over the Wigner degeneracy yields
\begin{align}
\sum_{n, \sigma}
u_{n} (\p,\sigma) \dual{u}_{n} (\p,\sigma)
&= b_{u} \slashed{p} + 2 m \mathds{1}
\, ,
\label{eq:sum_elko_u_12} \\
\sum_{n, \sigma}
v_{n} (\p,\sigma) \dual{v}_{n} (\p,\sigma)
&= b_{u} \slashed{p} -2 m \mathds{1}
\, .
\label{eq:sum_elko_v_12}
\end{align}
It is clear that the individual causality condition is identical to Eq.~\eqref{eq:causal_Dirac_b_1}, while
the causality mixing configuration of $b_{u, n}, b_{v, n}$ in Eq.~\eqref{eq:causal_Dirac_b_2} still leads to a non-zero causality mixing. 
As we proposed at the end of Section~\ref{sec:Dirac_0}, the Elko condition of Eq.~\eqref{eq:causal_elko_b_000} simplifies the ortho-normalization relations with Elko spinors in Eqs.~\eqref{eq:norm_elko_u_1}-\eqref{eq:norm_elko_uv_m_3}. These now take the form
\begin{align}
\dual{u}_{n}(\p,\sigma) u_{n'}(\p,\sigma') &= - \dual{v}_{n}(\p,\sigma) v_{n'}(\p,\sigma') = 2m \delta_{\sigma\sigma'} \delta_{n n'} \, ,
\label{eq:norm_elko_uv_2} \\
\dual{u}_{n}(\p,\sigma) v_{n'}(\p,\sigma) &= \dual{v}_{n}(\p,\sigma) u_{n'}(\p,\sigma) 
\nonumber \\
&= u^{\dag}_{n}(\p,\sigma) v_{n'}(-\p,\sigma) 
= v^{\dag}_{n}(\p,\sigma) u_{n'}(-\p,\sigma) = 0 \, ,
\label{eq:norm_elko_uv_m_12} \\
\dual{u}_{n}(\p,\sigma) v_{n'}(\p,-\sigma)  
&= \dual{v}_{n'}(\p,\sigma) u_{n}(\p,-\sigma)
\nonumber \\
&=u^{\dag}_{n}(\p,\sigma) v_{n'}(-\p,-\sigma) 
= v^{\dag}_{n'}(-\p,-\sigma) u_{n}(\p,\sigma)
\nonumber \\
&= 2m (-1)^{\frac{1}{2}+\sigma} \delta_{n n'} \, .
\label{eq:norm_elko_uv_m_22} 
\end{align}
Meanwhile, sums of Eqs.~\eqref{eq:sum_elko_u_1}-\eqref{eq:sum_elko_v_1} over the Wigner degeneracy and spin also take a simpler form
\begin{align}
\sum_{n, \sigma}
u_{n, \ell} (\p,\sigma) \dual{u}_{n, \ell'} (\p,\sigma)
&= 2 m \delta_{\ell \ell'}
\, ,
\label{eq:sum_elko_u_2} \\
\sum_{n, \sigma}
v_{n, \ell} (\p,\sigma) \dual{v}_{n, \ell'} (\p,\sigma)
&= -2 m \delta_{\ell \ell'}
\, .
\label{eq:sum_elko_v_2}
\end{align}

\subsection{Basis transformation \& causality mixing}
\label{sec:basisT_0}

Although the Wigner doublet is initially introduced as a common eigenstate of the Wigner degeneracy and spin in Section~\ref{sec:Spin_m_1}, one can generally redefine the basis of the one-particle states by mixing spin projections and the Wigner degeneracy. This basis redefinition can be written as
\begin{align}
|\p,\tau; \tilde{a}^{(c)} \rangle \equiv \sum_{n,\sigma} S^{(c)}_{(n,\sigma) \tau} |\p,\sigma, n; a^{(c)} \rangle \, ,
\label{eq:deg_op_ab2C_1}
\end{align}
where $S$ and $S^{(c)}$ are two unitary matrices corresponding to the Wigner-degenerate particle and its antiparticle, respectively, satisfying
\begin{align}
S^{(c)-1}_{\tau (n,\sigma)} = S^{(c)\dag}_{\tau (n,\sigma)} = S_{(n,\sigma) \tau}^{(c)*} \, ,
\label{eq:deg_ab2C_mat_u}
\end{align}
with $\tau=1,\cdots, 4$, and using the following identifications for the index of matrix elements
\begin{alignat}{2}
\tau &= 1 \equiv \left(+\frac{1}{2},+\frac{1}{2} \right) \, , \quad
&\tau &= 2 \equiv \left(+\frac{1}{2},-\frac{1}{2} \right) \, , \\
\tau &= 3 \equiv \left(-\frac{1}{2},+\frac{1}{2} \right) \, , \quad
&\tau &= 4 \equiv \left(-\frac{1}{2},-\frac{1}{2} \right) \, .
\end{alignat} 
Consequently, the creation and annihilation operators in the new basis are given by
\begin{alignat}{2}
\tilde{a}_{\tau}^{(c) \dag}(\p) &\equiv \sum_{n,\sigma} S_{(n,\sigma) \tau}^{(c)} a_{n}^{(c) \dag}(\p, \sigma) \, , \quad 
&a_{n}^{(c) \dag}(\p, \sigma) &= \sum_{\tau} S_{\tau (n,\sigma)}^{(c)-1} \tilde{a}_{\tau}^{(c) \dag}(\p) 
\, ,
\label{eq:deg_op_ab2C_1+} \\
\tilde{a}_{\tau}^{(c)}(\p) &\equiv \sum_{n,\sigma} S_{(n,\sigma) \tau}^{(c)*} a_{n}^{(c)}(\p, \sigma) \, ,
\quad 
&a_{n}^{(c)}(\p, \sigma) &= \sum_{\tau} S_{\tau (n,\sigma)}^{(c)-1*} \tilde{a}_{\tau}^{(c)}(\p) 
\, .
\label{eq:deg_op_ab2C_1-}
\end{alignat}
By imposing the unitarity condition on the matrices $S$ and $S^{c}$~\eqref{eq:deg_ab2C_mat_u}, the non-vanishing quantization relations in the new basis then take the canonical form
\begin{align}
\bigl\{ \tilde{a}_{\tau}(\p), \tilde{a}_{\tau'}^{\dag}(\p') \bigr\}
= \bigl\{ \tilde{a}_{\tau}^{c}(\p), \tilde{a}_{\tau'}^{c \dag}(\p') \bigr\}
= (2 \pi)^{3} \delta^{(3)}(\p-\p') \delta_{\tau \tau'}
\, .
\end{align}
The corresponding polarizations in the new basis are constructed so as to preserve the invariant sums appearing in the Wigner-degenerate fields~\eqref{eq:sf_W_20}
\begin{alignat}{2}
\sum_{\tau} \xi_{\tau}(\p) \tilde{a}_{\tau}(\p) &= \sum_{n, \sigma} u_{n}(\p,\sigma) a_{n}(\p,\sigma) \, , \quad 
&\sum_{\tau} \dual{\xi}_{\tau}(\p) \tilde{a}_{\tau}^{\dag}(\p) &= \sum_{n, \sigma} \dual{u}_{n}(\p,\sigma)a_{n}^{\dag}(\p,\sigma) \, ,
\\
\sum_{\tau} \zeta_{\tau}(\p) \tilde{a}_{\tau}^{c \dag}(\p) &= \sum_{n, \sigma} v_{n}(\p,\sigma) a_{n}^{c \dag}(\p,\sigma) \, , \quad 
&\sum_{\tau} \dual{\zeta}_{\tau}(\p) \tilde{a}_{\tau}^{c}(\p) &= \sum_{n, \sigma} \dual{v}_{n}(\p,\sigma)a^{c}_{n}(\p,\sigma) \, ,
\end{alignat}
so that the polarizations in the new basis are given by
\begin{alignat}{2}
\xi_{\tau}(\p) &= \sum_{n, \sigma} u_{n}(\p,\sigma) S_{\tau (n,\sigma)}^{-1*} ,  \ 
&\dual{\xi}_{\tau}(\p) &= \sum_{n, \sigma} \dual{u}_{n}(\p,\sigma) S_{\tau (n,\sigma)}^{-1} 
= \left[m^{-1} p_{\mu} \gamma^{\mu} \xi_{\tau}(\p) \right]^{\dagger} \gamma^{0} ,
\label{eq:elko_xi_0} \\
\zeta_{\tau}(\p) &= \sum_{n, \sigma} v_{n}(\p,\sigma) S_{\tau (n,\sigma)}^{c-1} ,  \  
&\dual{\zeta}_{\tau}(\p) &= \sum_{n, \sigma} \dual{v}_{n}(\p,\sigma) S_{\tau (n,\sigma)}^{c-1*} 
= - \left[m^{-1} p_{\mu} \gamma^{\mu} \zeta_{\tau}(\p) \right]^{\dagger} \gamma^{0} .
\label{eq:elko_zeta_0}
\end{alignat}
Let us now review the causality configuration from the perspective of basis redefinition.
Consider the initial basis of the two-fold Wigner-degenerate fields $\psi_{n}(x)$, $n=\pm \frac{1}{2}$~\eqref{eq:sf_W_20}, with sign factors in the non-zero causality mixing configuration~\eqref{eq:causal_Dirac_b_2}. 
We can then introduce the following basis transformation:
\begin{align}
S = \mathds{1}
\, , \quad 
S^{c} = \gamma^{0}
\, , 
\label{eq:cau_m2n_0}
\end{align}
which yields a new basis by flipping the Wigner degeneracy only for anti-particles
\begin{align}
\tilde{a}_{n}(\p,\sigma) &\equiv a_{n}(\p,\sigma)
\, , 
\label{eq:red_W_0} \\
\tilde{a}^{c \dag}_{n}(\p,\sigma) &\equiv \sum_{n'} \sigma^{1}_{n' n} a_{n'}^{c \dag}(\p,\sigma)
= \sum_{n',\sigma'} \gamma^{0}_{(n',\sigma') (n,\sigma)} a_{n'}^{c \dag}(\p,\sigma')
\, .
\label{eq:red_W_c_1}
\end{align}
It's remarkable that the new basis still consists of eigenstates of both the Wigner degeneracy and spin.
The associated polarizations are given by Eqs.~\eqref{eq:elko_xi_0}-\eqref{eq:elko_zeta_0}
\begin{align}
\xi_{n,\sigma}(\p) = u_{n}(\p,\sigma)
\, , \quad 
\zeta_{n,\sigma}(\p) = v_{-n}(\p,\sigma)
\, .
\end{align}
Thus, the new two-fold Wigner-degenerate fields take the form:
\begin{align}
\chi_{n}(x) &= \chi_{n}^{+}(x) + \chi_{n}^{-c}(x)
\, , 
\end{align}
with
\begin{align}
\chi_{n}^{+}(x)
&= \psi_{n}^{+}(x)
=
\int\dfrac{d^{3}p}{(2\pi)^{3}}\frac{1}{\sqrt{2E_{\p}}}\sum_{\sigma} e^{-ip\cdot x}u_{n}(\p,\sigma)\tilde{a}_{n}(\p,\sigma) 
\, , \\
\chi_{n}^{-c}(x)
&= \psi_{-n}^{-c}(x)
=
\int\dfrac{d^{3}p}{(2\pi)^{3}}\frac{1}{\sqrt{2E_{\p}}}\sum_{\sigma} e^{ip\cdot x} v_{-n}(\p,\sigma)\tilde{a}^{c \dag}_{n}(\p,\sigma) 
\, .
\end{align}
The Wigner superposition field $\lambda(x)$~\eqref{eq:field_1+1_W_0} remains invariant under this basis transformation, and can now be rewritten as
\begin{align}
\lambda (x)
= \dfrac{1}{\sqrt{2}} \left[ \chi_{+\frac{1}{2}}(x) + \chi_{-\frac{1}{2}}(x) \right]
\, .
\end{align}
In this new basis, the causality conditions~\eqref{eq:casual_1}-\eqref{eq:casual_1+1_elko_0} can now be satisfied individually by $\chi_{+\frac{1}{2}}(x)$ and $\chi_{-\frac{1}{2}}(x)$, i.e., causality non-mixing is required. 
Moreover, by taking the reversed basis transformation, one can realize the causality mixing configuration if it was initially absent, satisfying the Elko condition~\eqref{eq:causal_elko_b_000}. In fact, to construct a self-consistent QFT for the Wigner superposition field $\lambda (x)$~\eqref{eq:field_1+1_W_0} in the canonical quantization formalism, the Elko condition~\eqref{eq:causal_elko_b_000} is mandatory to impose. This fundamental requirement will be carefully demonstrated in Section~\ref{sec:can_W_s_1}.
Thus, causality mixing is not a physical property of the Wigner-degenerate fields, but an artifact consequence of the chosen basis.
With a suitable basis transformation, one can freely switch between causality mixing and non-mixing scenarios.
Nevertheless, for practical purposes, one can adopt a specific basis for derivations, where causality mixing serves as a useful tool for classifying distinct field configurations.


\section{$P$, $T$, and $C$ on the Wigner superposition field}\label{sec:pct}

The continuous Lorentz transformations act trivially on the Wigner degeneracy, implying that the Wigner doublet furnishes a reducible representation. 
Therefore, to promote the Wigner doublet into a simple yet irreducible representation of the
extended Poincar\'{e} group, one must require that the discrete inversions mix the Wigner (anti-)doublets.

\subsection{Two spacetime inversions}

The actions of the parity (space inversion) $P$ and the time-reversal $T$ on the Wigner-degenerate states are given by
\begin{align}
    P|\p,\sigma,n; a^{(c)}\rangle&=\sum_{n'}D^{(c)}_{n'n}(\mathscr{P})|-\p,\sigma,n'; a^{(c)}\rangle \, ,
    \label{eq:P_state_W_ab}\\
    T|\p,\sigma,n; a^{(c)}\rangle&=(-1)^{\frac{1}{2}-\sigma}\sum_{n'}D_{n'n}^{(c)}(\mathscr{T})|-\p,-\sigma,n'; a^{(c)}\rangle \, .
    \label{eq:T_state_W_ab}
\end{align}
Since $P$ is linear and unitary, its associated matrix $D^{(c)}(\mathscr{P})$ is a $2\times2$ unitary matrix, which can always be diagonalized in a suitable basis. 
In contrast, $T$ is antiunitary and antilinear. We can only block-diagonalize the time-reversal matrix $D^{(c)}(\mathscr{T})$ in general.
We assume the existence of a basis in which the Wigner degeneracy is manifest, and the corresponding transformation matrices for $P$ and $T$ are diagonal and anti-diagonal respectively
\begin{align}
D(\mathscr{P})&=\left[\begin{matrix}
        \eta_{+\frac{1}{2}} & 0 \\
        0 & \eta_{-\frac{1}{2}}
    \end{matrix}\right] \, ,
    \quad \quad 
    D^{c}(\mathscr{P})=\left[\begin{matrix}
        \eta^{c}_{+\frac{1}{2}} & 0 \\
        0 & \eta^{c}_{-\frac{1}{2}}
    \end{matrix}\right] \, , 
    \label{mat:P_state_W_ab_2}
    \\
    D(\mathscr{T})&=\left[\begin{matrix}
        0 & e^{i\phi/2} \\
        e^{-i\phi/2} & 0
    \end{matrix}\right] \, ,
    \quad
    D^{c}(\mathscr{T})=\left[\begin{matrix}
        0 & e^{i\phi^{c}/2} \\
        e^{-i\phi^{c}/2} & 0
    \end{matrix}\right] \, , 
    \label{mat:T_state_W_ab_2}
\end{align}
where phases $\eta_{\pm \frac{1}{2}}$ and $\eta^{c}_{\pm \frac{1}{2}}$ are intrinsic parities for the particle and the anti-particle states with the Wigner degeneracies $n=\pm\frac{1}{2}$ respectively. $e^{i\frac{\phi}{2}}$ and $e^{i\frac{\phi^{c}}{2}}$ are the time-reversal phases for the Wigner-degenerate particles and the anti-particles.
To make the notation clear, we denote the $2 \times 2$ matrix element with the Wigner degeneracy index $n=\pm\frac{1}{2}$ as 
\begin{equation}
    D=\left[\begin{matrix}
        D_{+\frac{1}{2},+\frac{1}{2}} & D_{+\frac{1}{2},-\frac{1}{2}} \\
        D_{-\frac{1}{2},+\frac{1}{2}} & D_{-\frac{1}{2},-\frac{1}{2}}
    \end{matrix}\right]\,.
\end{equation}
The diagonal structure of the parity matrix $D^{(c)}(\mathscr{P})$~\eqref{mat:P_state_W_ab_2} indicates that $P$ preserves the Wigner degeneracy. In contrast, the anti-diagonal time-reversal matrix $D^{(c)}(\mathscr{T})$~\eqref{mat:T_state_W_ab_2} results in an exchange of the Wigner degeneracy under the time-reversal $T$.
Two inversions on the annihilation and creation operators can be derived by applying Eqs.~\eqref{eq:P_state_W_ab}-\eqref{eq:T_state_W_ab} to Eq.~\eqref{eq:deg_op_ab} with the corresponding transformation matrices $D^{(c)}_{n'n}(\mathscr{P})$~\eqref{mat:P_state_W_ab_2} and $D^{(c)}_{n'n}(\mathscr{T})$~\eqref{mat:T_state_W_ab_2}:
\begin{align}
P a_{n}^{(c)\dag}(\p, \sigma)  P^{-1} &= 
\eta^{(c)}_{n} a_{n}^{(c)\dag}(-\p, \sigma) \, ,
\label{eq:P_part_W_1} \\
P a_{n}^{(c)}(\p, \sigma)  P^{-1} &= 
\eta^{(c)*}_{n} a_{n}^{(c)}(-\p, \sigma) \, ,
\label{eq:P_part-_W_1}
\\
T a_{n}^{(c)\dag}(\p, \sigma)  T^{-1} &= 
(-1)^{\frac{1}{2}-\sigma} e^{-i n\phi^{(c)}} a_{-n}^{(c)\dag}(-\p, -\sigma) \, ,
\label{eq:T_part_W_1} \\
T a_{n}^{(c)}(\p, \sigma)  T^{-1} &= 
(-1)^{\frac{1}{2}-\sigma} e^{i n\phi^{(c)}} a_{-n}^{(c)}(-\p, -\sigma) \, .
\label{eq:T_part-_W_1}
\end{align}
Applying these results to the Wigner-degenerate fields $\psi_{n}(x)$, $n=\pm\frac{1}{2}$~\eqref{eq:sf_W_20}, we find that
\begin{align}
P\psi_{n}(x) P^{-1}
&=\gamma^{0} \int\frac{d^{3}p}{(2\pi)^{3}}\frac{1}{\sqrt{2E_{\p}}}\sum_{\sigma}
\biggl[\eta^{*}_{n} b_{u, n} e^{-ip\cdot \mathscr{P} x} u_{n}(\p,\sigma)a_{n}(\p,\sigma) 
\nonumber\\
&\quad \quad \quad \quad \quad \quad \quad \quad \quad \quad \quad 
+ \eta^{c}_{n} b_{v, n} e^{ip\cdot \mathscr{P} x} v_{n}(\p,\sigma)a^{c \dag}_{n}(\p,\sigma)\biggr] 
\, , 
\label{eq:P_W_field_1} \\
T\psi_{n}(x) T^{-1}
&=\gamma^{1} \gamma^{3} \left( P_{L} + \Delta P_{R} \right)
\nonumber\\
&\quad \quad \  \times \int\frac{d^{3}p}{(2\pi)^{3}}\frac{1}{\sqrt{2E_{\p}}}\sum_{\sigma}
\biggl[e^{i n \phi}e^{-ip\cdot \mathscr{T}x} u_{-n}(\p,\sigma) a_{-n}(\p,\sigma) 
\nonumber\\
&\quad \quad \quad \quad \quad \quad \quad \quad \quad \quad \quad 
+ e^{-i n\phi^{c}}e^{ip\cdot \mathscr{T}x} v_{-n}(\p,\sigma) a^{c \dag}_{-n}(\p,\sigma) \biggr] 
\label{eq:T_W_field_1} \, ,
\end{align}
where we have inserted the spinor identities in Eqs.~\eqref{eq:id_Dirac_uv_01}-\eqref{eq:id_Dirac_uv_03} and the causality condition~\eqref{eq:causal_Dirac_b_00}.
If we expect the two inversions to map the Wigner superposition field $\lambda (x)$~\eqref{eq:field_1+1_W_0} at some point $x$ into something proportional to itself at the corresponding point $x'$, it is necessary to require
\begin{align}
\hat{\eta}_{P} &\equiv
\eta_{+\frac{1}{2}} b_{u, +\frac{1}{2}} = \eta^{c*}_{+\frac{1}{2}} b_{v, +\frac{1}{2}}
=
\eta_{-\frac{1}{2}} b_{u, -\frac{1}{2}} = \eta^{c*}_{-\frac{1}{2}} b_{v, -\frac{1}{2}}
\, ,
\label{eq:P_W_phi_s} \\
\hat{\eta}_{T} &\equiv e^{i\phi/2} = e^{-i\phi^{c}/2} = \pm 1 \, ,
\label{eq:T_W_phi_s}
\end{align}
which provide stronger constraints on the intrinsic parities of Wigner-degenerate particles and their anti-particles than those derived in the doublet framework~\cite{Lee:2025kpz}. 
Furthermore, Eq.~\eqref{eq:P_W_phi_s} suggests an additional possibility that 
if we impose the  causality mixing configuration given in Eq.~\eqref{eq:causal_Dirac_b_2},
the intrinsic parities of an Elko particle and its anti-particle become anomalously related as
\begin{align}
\eta^{*}_{n} = \eta^{c}_{n} = -\eta^{*}_{-n} = -\eta^{c}_{-n} \, ,
\label{eq:P_W_phi_s_2} 
\end{align}
where, notably, the expected relative minus sign is absent in the first and the third equality. Additionally, the second equality introduces an extra relation that does not appear in the doublet framework.
Imposing Eqs.~\eqref{eq:P_W_phi_s}-\eqref{eq:T_W_phi_s} to Eqs.~\eqref{eq:P_W_field_1}-\eqref{eq:T_W_field_1}, we obtain the $P$, $T$ transformations on the Wigner superposition field $\lambda (x)$~\eqref{eq:field_1+1_W_0}
\begin{align}
P\lambda (x) P^{-1} &=
\hat{\eta}_{P}^{*} \gamma^{0} \lambda (\mathscr{P} x) \, , 
\label{eq:P_W_field_s_0} \\
T\lambda (x) T^{-1} &=
\hat{\eta}_{T} \gamma^{1}\gamma^{3} \left( P_{L} + \Delta P_{R} \right) \lambda (\mathscr{T}x) \, ,
\label{eq:T_W_field_s_0}
\end{align}
along with the corresponding actions on its Dirac dual field $\bar{\lambda} (x)$~\eqref{eq:Dirac_dual_1}
\begin{align}
P \bar{\lambda}(x) P^{-1} &=
\hat{\eta}_{P} \bar{\lambda}(\mathscr{P} x) \gamma^{0} \, , 
\label{eq:P_W_field_s_Dirac_0} \\
T \bar{\lambda}(x) T^{-1} &=
\hat{\eta}_{T} \bar{\lambda}(\mathscr{T}x) \left( P_{R} + \Delta P_{L} \right) \gamma^{3}\gamma^{1} \, ,
\label{eq:T_W_field_s_Dirac_0}
\end{align}
and on its Elko dual field $\dual{\lambda} (x)$~\eqref{eq:elko_dual_0}
\begin{align}
P \dual{\lambda}(x) P^{-1} &=
\hat{\eta}_{P} \dual{\lambda}(\mathscr{P} x) \gamma^{0} \, , 
\label{eq:P_W_field_s_elko_0} \\
T \dual{\lambda}(x) T^{-1} &=
\hat{\eta}_{T} \dual{\lambda}(\mathscr{T}x) \left( P_{L} + \Delta P_{R} \right) \gamma^{3}\gamma^{1} \, .
\label{eq:T_W_field_s_elko_0}
\end{align}
The general formula of the two discrete inversions on the Wigner doublets~\eqref{eq:P_state_W_ab}-\eqref{eq:T_state_W_ab} can be simplified via Eqs.~\eqref{mat:P_state_W_ab_2}-\eqref{mat:T_state_W_ab_2} and the consistency conditions~\eqref{eq:P_W_phi_s}-\eqref{eq:T_W_phi_s}
\begin{align}
    P|\p,\sigma,n; a\rangle&= b_{u, n} \hat{\eta}_{P} |-\p,\sigma,n; a\rangle \, ,
    \label{eq:P_state_W_a_2} \\
    P|\p,\sigma,n; a^{c}\rangle&= b_{v, n} \hat{\eta}_{P}^{*} |-\p,\sigma,n; a^{c}\rangle \, ,
    \label{eq:P_state_W_b_2}\\
    T|\p,\sigma,n; a\rangle&=(-1)^{\frac{1}{2}-\sigma} \hat{\eta}_{T} |-\p,-\sigma,-n; a\rangle \, ,
    \label{eq:T_state_W_a_2} \\
    T|\p,\sigma,n; a^{c}\rangle&=(-1)^{\frac{1}{2}-\sigma} \hat{\eta}_{T} |-\p,-\sigma,-n; a^{c}\rangle \, .
    \label{eq:T_state_W_b_2}
\end{align}
In particular, there may exist an additional internal symmetry unitary operator $\hat{S}_{T}$ that acts non-trivially on the Wigner doublet as
\begin{align}
\hat{S}_{T} |\p,\sigma,n; a\rangle = \hat{\eta}_{T} |\p,\sigma,-n; a\rangle \, , 
\quad
\hat{S}_{T} |\p,\sigma,n; a^{c}\rangle = \hat{\eta}_{T} |\p,\sigma,-n; a^{c}\rangle \, , 
\label{eq:S_re_T_0}
\end{align}
but trivially on usual particles
\begin{align}
\hat{S}_{T} |\p,\sigma; a^{(c)}\rangle &= \eta_{S} |\p,\sigma; a^{(c)}\rangle \, , 
\ \ \text{with} \ \ 
\eta_{S} = \pm 1 \, ,
\end{align}
so that $\hat{S}_{T}=\hat{S}_{T}^{-1}=\hat{S}_{T}^{\dag}$.
Applying $\hat{S}_{T}$ to the Wigner-degenerate fields $\psi_{n}(x)$, $n=\pm\frac{1}{2}$~\eqref{eq:sf_W_20} yields
\begin{align}
\hat{S}_{T} \psi_{n}(x) \hat{S}_{T}^{-1}
= \hat{\eta}_{T} \left( P_{L} + \Delta P_{R} \right) \psi_{-n}(x) 
\, , 
\label{eq:S_re_T_field_s1}
\end{align}
where we have inserted the spinor identity in Eq.~\eqref{eq:id_Dirac_uv_03} and the causality condition~\eqref{eq:causal_Dirac_b_00}.
As a result, the Wigner superposition field $\lambda (x)$~\eqref{eq:field_1+1_W_0} transforms under $\hat{S}_{T}$ 
\begin{align}
\hat{S}_{T} \lambda (x) \hat{S}_{T}^{-1} = \hat{\eta}_{T} \left( P_{L} + \Delta P_{R} \right) \lambda (x) \, ,
\label{eq:S_re_T_field_s2}
\end{align}
with the corresponding transformations on its Dirac dual field $\bar{\lambda} (x)$~\eqref{eq:Dirac_dual_1} and Elko dual field $\dual{\lambda} (x)$~\eqref{eq:elko_dual_0}
\begin{align}
\hat{S}_{T} \bar{\lambda}(x) \hat{S}_{T}^{-1} &=
\hat{\eta}_{T} \bar{\lambda}(x) \left( P_{R} + \Delta P_{L} \right) \, ,
\label{eq:T_W_field_s_Dirac_s0}
\\
\hat{S}_{T} \dual{\lambda}(x) \hat{S}_{T}^{-1} &=
\hat{\eta}_{T} \dual{\lambda}(x) \left( P_{L} + \Delta P_{R} \right) \, .
\label{eq:T_W_field_s_elko_s0}
\end{align}
One can then redefine the time-reversal operator as
\begin{align}
T' \equiv \hat{S}_{T}^{-1} T \, ,
\end{align}
which transforms all particles in the conventional pattern
\begin{align}
T' |\p,\sigma,n; a^{(c)}\rangle&=(-1)^{\frac{1}{2}-\sigma} |-\p,-\sigma,n; a^{(c)} \rangle \, , \\
T' |\p,\sigma; a^{(c)}\rangle &= (-1)^{j-\sigma} \eta_{S}^{(*)} |-\p, -\sigma; a^{(c)}\rangle \, .
\end{align}
Thus, for a physically nontrivial $T$ invariant Wigner theory of time-reversal doublets, such an internal symmetry $\hat{S}_{T}$ must be absent. This requirement is not only essential for the physical relevance of the Wigner degeneracy but also resolves ambiguities in defining spacetime reflection operators, as discussed in Ref.~\cite{Lee:2025kpz,Lee:1966ik,daSilvaBarbosa:2023xfy}. 
We will further investigate the role of this internal symmetry $\hat{S}_{T}$ within the context of the free theory in Section~\ref{sec:elko_KG_1}.

\subsection{Charge-conjugation}

After investigating the $P$ and $T$ transformations, we now introduce the charge-conjugation symmetry, which relates particles to their corresponding antiparticles. Unlike the isometry group of the Minkowski spacetime, charge-conjugation is an internal symmetry, independent of the spacetime structure. Following the methodology outlined in our previous work~\cite{Lee:2025kpz}, we consider the unitary charge-conjugation operator $C $, which maps a Wigner-degenerate particle state $|\p,\sigma,n;a\rangle$ to a superposition of anti-particle states $|\p,\sigma,n;a^{c}\rangle$, summed over the degenerate degrees of freedom:
\begin{equation}
    C |\p,\sigma,n;a\rangle=\sum_{n'}D_{n'n}(\mathscr{C})|\p,\sigma,n';a^{c}\rangle \, ,
    \label{eq:C_state_W_ab_1}
\end{equation}
and conversely, transforms $|\p,\sigma,n;a^{c}\rangle$ into a superposition of $|\p,\sigma,n;a\rangle$
\begin{equation}
    C |\p,\sigma,n;a^{c}\rangle=\sum_{n'}D^{c}_{n'n}(\mathscr{C})|\p,\sigma,n';a\rangle \,.
    \label{eq:C_state_W_ab_2}
\end{equation}
Therefore, under charge-conjugation, the annihilation and creation operators transform as
\begin{alignat}{2}
    C a^{\dag}_{n}(\p,\sigma) C^{-1} &=\sum_{n'}D_{n'n}(C)a^{c\dag}_{n'}(\p,\sigma)\, ,
    \quad
    &C a_{n}(\p,\sigma) C^{-1} &=\sum_{n'}D^{*}_{n'n}(C)a^{c}_{n'}(\p,\sigma)\, ,
    \label{eq:C_part_W_1+} \\
    C a^{c\dag}_{n}(\p,\sigma) C^{-1} &=\sum_{n'}D^{c}_{n'n}(C)a^{\dag}_{n'}(\p,\sigma)\, ,
    \quad
    &C a^{c}_{n}(\p,\sigma) C^{-1} &=\sum_{n'}D^{c*}_{n'n}(C)a_{n'}(\p,\sigma)\, ,
    \label{eq:C_part_W_1c}
\end{alignat}
which induce the charge-conjugation transformation for the Wigner-degenerate fields $\psi_{n}(x)$, $n=\pm\frac{1}{2}$~\eqref{eq:sf_W_20}
\begin{align}
C \psi_{n}(x) C^{-1} 
&=-i\gamma^{2}\int\frac{d^{3}p}{(2\pi)^{3}}\frac{1}{\sqrt{2E_{\p}}}
\nonumber\\
&\quad   \times \sum_{\sigma,n'}
\biggl[e^{-ip\cdot x} \left( b_{u,n} P_{L} - b_{v,\pm n} P_{R} \right) v_{\pm n}^{*}(\p,\sigma) D^{*}_{n'n}(\mathscr{C}) a^{c}_{n'}(\p,\sigma) 
\nonumber\\
&\quad \quad \quad  
+ e^{ip\cdot x} \left( b_{u,\pm n} P_{R} - b_{v,n} P_{L} \right) u_{\pm n}^{*}(\p,\sigma)D^{c}_{n'n}(\mathscr{C}) a^{\dag}_{n'}(\p,\sigma) \biggr] 
\, ,
\label{eq:C_W_field_1} 
\end{align}
or
\begin{align}
C \psi_{n}(x) C^{-1} 
&=-i\gamma^{2}\int\frac{d^{3}p}{(2\pi)^{3}}\frac{1}{\sqrt{2E_{\p}}}
\nonumber\\
&\quad   \times \sum_{\sigma,n'}
\biggl[e^{-ip\cdot x} \left( b_{u,n} P_{L} - b_{v,\pm n} P_{R} \right) v_{\pm n}^{*}(\p,\sigma) D^{*}_{n'n}(\mathscr{C}) a^{c}_{n'}(\p,\sigma) 
\nonumber\\
&\quad \quad \quad  
+ e^{ip\cdot x} \left( b_{u,\mp n} P_{R} - b_{v,n} P_{L} \right) u_{\mp n}^{*}(\p,\sigma)D^{c}_{n'n}(\mathscr{C}) a^{\dag}_{n'}(\p,\sigma) \biggr] 
\, ,
\label{eq:C_W_field_2} 
\end{align}
where we have inserted the spinor identities in Eqs.~\eqref{eq:id_Dirac_u_04}-\eqref{eq:id_Dirac_v_04}.
In fact, as demonstrated in Section~\ref{sec:can_W_s_1}, to construct a self-consistent QFT for the Wigner superposition field $\lambda (x)$~\eqref{eq:field_1+1_W_0} in the canonical quantization formalism, we must impose the Elko condition on the sign factors $b_{u,n}$ and $b_{v,n}$ in Eq.~\eqref{eq:causal_elko_b_000}, which means we must treat it as an Elko field. As a consequence, the charge-conjugation transformations in Eqs.~\eqref{eq:C_W_field_1}-\eqref{eq:C_W_field_2} respectively reduce to
\begin{align}
C \psi_{n}(x) C^{-1} 
&=-i\gamma^{2}\int\frac{d^{3}p}{(2\pi)^{3}}\frac{1}{2\sqrt{E_{\p}}}
\nonumber\\
&\quad   \times \sum_{\sigma,n'}
\biggl[e^{-ip\cdot x} \left( b_{u,n} P_{L} \mp b_{v,n} P_{R} \right) v_{\pm n}^{*}(\p,\sigma) D^{*}_{n'n}(\mathscr{C}) a^{c}_{n'}(\p,\sigma) 
\nonumber\\
&\quad \quad \quad \, 
+ e^{ip\cdot x} \left( \pm b_{u,n} P_{R} - b_{v,n} P_{L} \right) u_{\pm n}^{*}(\p,\sigma)D^{c}_{n'n}(\mathscr{C}) a^{\dag}_{n'}(\p,\sigma) \biggr] 
\, ,
\label{eq:C_W_field_12} 
\end{align}
and 
\begin{align}
C \psi_{n}(x) C^{-1} 
&=-i\gamma^{2}\int\frac{d^{3}p}{(2\pi)^{3}}\frac{1}{2\sqrt{E_{\p}}}
\nonumber\\
&\quad   \times \sum_{\sigma,n'}
\biggl[e^{-ip\cdot x} \left( b_{u,n} P_{L} \mp b_{v,n} P_{R} \right) v_{\pm n}^{*}(\p,\sigma) D^{*}_{n'n}(\mathscr{C}) a^{c}_{n'}(\p,\sigma) 
\nonumber\\
&\quad \quad \quad \, 
+ e^{ip\cdot x} \left( \mp b_{u,n} P_{R} - b_{v,n} P_{L} \right) u_{\mp n}^{*}(\p,\sigma)D^{c}_{n'n}(\mathscr{C}) a^{\dag}_{n'}(\p,\sigma) \biggr] 
\, .
\label{eq:C_W_field_22} 
\end{align}
Then, we need to consider the two distinct causality cases separately:
\begin{itemize}
\item Non-mixing: $b_{n}\equiv b_{u,n}=-b_{v,n}=-b_{-n}$ required in Eq.~\eqref{eq:causal_Dirac_b_1}. 
Under this condition, the charge-conjugation transformations of the Wigner-degenerate fields $\psi_{n}(x)$ in Eqs.~\eqref{eq:C_W_field_12}-\eqref{eq:C_W_field_22} then converge to a uniform result:
\begin{align}
C \psi_{n,\ell} (x) C^{-1} 
&= \sum_{\ell'} \left[ \mathcal{C} \left( P_{L} \pm P_{R} \right) \right]_{\ell \ell'} \int\frac{d^{3}p}{(2\pi)^{3}}\frac{1}{2\sqrt{E_{\p}}} 
\nonumber\\
&\quad \quad \quad \    \times \sum_{\sigma,n'}  
\biggl[e^{-ip\cdot x} \dual{v}_{\pm n, \ell'}(\p,\sigma) D^{*}_{n'n}(\mathscr{C}) a^{c}_{n'}(\p,\sigma) 
\nonumber\\
&\quad \quad \quad \quad \quad \quad \,  
+ e^{ip\cdot x} \dual{u}_{\pm n, \ell'}(\p,\sigma)D^{c}_{n'n}(\mathscr{C}) a^{\dag}_{n'}(\p,\sigma) \biggr] 
\, .
\label{eq:C_W_field_c_1} 
\end{align}
If we expect charge conjugation to map $\lambda (x)$ to a linear transformation of $\dual{\lambda}(x)$, we must have one of the following cases 
\begin{enumerate}[label=\arabic*)]  
\item $D^{*}_{n'n}(\mathscr{C})=D^{c}_{n'n}(\mathscr{C}) = \hat{\eta}_{C} \delta_{n'n}$, leading to
\begin{align}
C \lambda_{\ell} (x) C^{-1} 
&= \hat{\eta}_{C} \sum_{\ell'} \mathcal{C}_{\ell \ell'} \dual{\lambda}_{\ell'}(x)
\, ,
\label{eq:C_W_field_s_11} 
\\
C \dual{\lambda}_{\ell} (x) C^{-1} 
&= \hat{\eta}_{C}^{*} \sum_{\ell'} \lambda_{\ell'}(x) \mathcal{C}_{\ell' \ell} 
\, ,
\label{eq:C_W_field_s_elko_11} 
\end{align}
\item $D^{*}_{n'n}(\mathscr{C})=D^{c}_{n'n}(\mathscr{C}) = \hat{\eta}_{C} \delta_{n'-n}$, leading to
\begin{align}
C \lambda_{\ell} (x) C^{-1} 
&=-\hat{\eta}_{C} \sum_{\ell'} \left(\mathcal{C} \gamma^{5}\right)_{\ell \ell'} \dual{\lambda}_{\ell'}(x)
\, ,
\label{eq:C_W_field_s_12} 
\\
C \dual{\lambda}_{\ell} (x) C^{-1} 
&= -\hat{\eta}_{C}^{*} \sum_{\ell'} \lambda_{\ell'}(x) \left(\mathcal{C} \gamma^{5}\right)_{\ell' \ell} 
\, ,
\label{eq:C_W_field_s_elko_12} 
\end{align}
\end{enumerate}
where $\hat{\eta}_{C}$ is a charge-conjugation phase factor.
\item Mixing: $b_{u,n}=b_{v,n}=-b_{u,-n}=-b_{v,n}$ required in Eq.~\eqref{eq:causal_Dirac_b_2}. 
The charge-conjugation transformation of the Wigner-degenerate fields $\psi_{n}(x)$ in Eqs.~\eqref{eq:C_W_field_12}-\eqref{eq:C_W_field_22} then converge to
\begin{align}
C \psi_{n,\ell} (x) C^{-1} 
&= \sum_{\ell'} \left[ \mathcal{C} \left( P_{L} \mp P_{R} \right) \right]_{\ell \ell'} \int\frac{d^{3}p}{(2\pi)^{3}}\frac{1}{2\sqrt{E_{\p}}} 
\nonumber\\
&\quad \quad \quad \    \times \sum_{\sigma,n'}  
\biggl[e^{-ip\cdot x} \dual{v}_{\pm n, \ell'}(\p,\sigma) D^{*}_{n'n}(\mathscr{C}) a^{c}_{n'}(\p,\sigma) 
\nonumber\\
&\quad \quad \quad \quad \quad \quad \,  
+ e^{ip\cdot x} \dual{u}_{\pm n, \ell'}(\p,\sigma)D^{c}_{n'n}(\mathscr{C}) a^{\dag}_{n'}(\p,\sigma) \biggr] 
\, .
\label{eq:C_W_field_c_2} 
\end{align}
Following the same methodology proposed for the causality non-mixing case in Eqs.~\eqref{eq:C_W_field_s_11}-\eqref{eq:C_W_field_s_12}, if the charge conjugation transforms $\lambda (x)$ to something proportional to $\dual{\lambda}(x)$, we have the following two possible transformations: 
\begin{enumerate}[label=\arabic*)]  
\item $D^{*}_{n'n}(\mathscr{C})=D^{c}_{n'n}(\mathscr{C}) = \hat{\eta}_{C} \delta_{n'n}$, leading to
\begin{align}
C \lambda_{\ell} (x) C^{-1} 
&= -\hat{\eta}_{C} \sum_{\ell'} \left(\mathcal{C} \gamma^{5}\right)_{\ell \ell'} \dual{\lambda}_{\ell'}(x)
\, ,
\label{eq:C_W_field_s_21} 
\\
C \dual{\lambda}_{\ell} (x) C^{-1} 
&= -\hat{\eta}_{C}^{*} \sum_{\ell'} \lambda_{\ell'}(x) \left(\mathcal{C} \gamma^{5}\right)_{\ell' \ell} 
\, .
\label{eq:C_W_field_s_elko_21} 
\end{align}
\item $D^{*}_{n'n}(\mathscr{C})=D^{c}_{n'n}(\mathscr{C}) = \hat{\eta}_{C} \delta_{n'-n}$, leading to
\begin{align}
C \lambda_{\ell} (x) C^{-1} 
&=\hat{\eta}_{C} \sum_{\ell'} \mathcal{C}_{\ell \ell'} \dual{\lambda}_{\ell'}(x)
\, ,
\label{eq:C_W_field_s_22} 
\\
C \dual{\lambda}_{\ell} (x) C^{-1} 
&= \hat{\eta}_{C}^{*} \sum_{\ell'} \lambda_{\ell'}(x) \mathcal{C}_{\ell' \ell} 
\, ,
\label{eq:C_W_field_s_elko_22} 
\end{align}
\end{enumerate}
where $\hat{\eta}_{C}$ is again a charge-conjugation phase factor.   
\end{itemize}
We can see that both of the two causality cases impose the same constraint on the transformation matrices of the charge-conjugation
\begin{align}
D^{*}_{n'n}(\mathscr{C})=D^{c}_{n'n}(\mathscr{C}) = \hat{\eta}_{C} \delta_{n' \pm n} 
\, .
\label{eq:C_W_field_s_con} 
\end{align}
However, the transformation results are dramatically interchanged. In particular, the transformations in Eqs.~\eqref{eq:C_W_field_s_11}-\eqref{eq:C_W_field_s_elko_12} for the causality non-mixing case are swapped with those in Eqs.~\eqref{eq:C_W_field_s_21}-\eqref{eq:C_W_field_s_elko_22} for the causality mixing case. This interchange can be understood through the basis transformation presented in Eqs.~\eqref{eq:cau_m2n_0}-\eqref{eq:red_W_c_1}, which effectively exchanges the cases with and without causality mixing.
The charge-conjugation transformation on the creation and annihilation operators in the new basis can be derived from their counterparts in the initial basis, as given by Eqs.~\eqref{eq:C_part_W_1+}-\eqref{eq:C_part_W_1c}:
\begin{align}
    C \tilde{a}^{\dag}_{n}(\p,\sigma) C^{-1} &=\sum_{n'}D_{-n'n}(\mathscr{C})\tilde{a}^{c\dag}_{n'}(\p,\sigma)
    =\sum_{n'}\tilde{D}_{n'n}(\mathscr{C})\tilde{a}^{c\dag}_{n'}(\p,\sigma) \, ,\\
    C \tilde{a}_{n}(\p,\sigma) C^{-1} &=\sum_{n'}D^{*}_{-n'n}(\mathscr{C})\tilde{a}^{c}_{n'}(\p,\sigma)
    =\sum_{n'}\tilde{D}^{*}_{n'n}(\mathscr{C})\tilde{a}^{c}_{n'}(\p,\sigma) \, ,\\
    C \tilde{a}^{c\dag}_{n}(\p,\sigma) C^{-1} &=\sum_{n'}D^{c}_{n'-n}(\mathscr{C})\tilde{a}^{\dag}_{n'}(\p,\sigma)
    =\sum_{n'}\tilde{D}^{c}_{n'n}(\mathscr{C})\tilde{a}^{\dag}_{n'}(\p,\sigma) \, ,\\
    C \tilde{a}^{c}_{n}(\p,\sigma) C^{-1} &=\sum_{n'}D^{c*}_{n'-n}(\mathscr{C})\tilde{a}_{n'}(\p,\sigma)
    =\sum_{n'}\tilde{D}^{c*}_{n'n}(\mathscr{C})\tilde{a}_{n'}(\p,\sigma) \, ,
\end{align}
which leads to the relationship between the charge-conjugation matrices of the initial basis and the new basis:
\begin{align}
D_{-n'n}(\mathscr{C}) =  \tilde{D}_{n'n}(\mathscr{C})
\, , \quad 
D^{c}_{n'-n}(\mathscr{C}) = \tilde{D}^{c}_{n'n}(\mathscr{C})
\, .
\end{align}
More explicitly, $D^{*}_{n'n}(\mathscr{C})=D^{c}_{n'n}(\mathscr{C}) = \hat{\eta}_{C} \delta_{n'\pm n}$ leads to $\tilde{D}^{*}_{n'n}(\mathscr{C})=\tilde{D}^{c}_{n'n}(\mathscr{C}) = \hat{\eta}_{C} \delta_{n' \mp n}$, and vice versa. 
This implies that the charge-conjugation results of the Elko field must correspond to an exchanged configuration of the transformation matrices in the two causality scenarios.

\section{Canonical quantization of the Wigner superposition field}
\label{sec:can_W_s_1}


In modern QFT, the canonical formalism, built on defining the Lagrangian and implementing canonical quantization, forms the fundamental framework for analyzing physical systems. This approach offers a systematic way to identify symmetries, including Lorentz (or Poincar\'e) invariance and other imposed symmetries.
To fully capture the physical nature of the two-fold Wigner degeneracy, we have developed a Lagrangian and canonical quantization formalism for the doublet framework in our previous work~\cite{Lee:2025kpz}.
In this section, following the same methodology, we will establish a suitable canonical quantization formalism for the Wigner superposition spinor field.

We start with the fact that the polarizations are simply eigenstates of $p_{\mu} \gamma^{\mu}$ as shown in Eqs.~\eqref{eq:Dirac_eq_u_gen_1}-\eqref{eq:Dirac_eq_v_gen_1}. This suggests that the Wigner-degenerate fields $\psi_{n}(x)$, $n=\pm \frac{1}{2}$~\eqref{eq:sf_W_20} can potentially be made to satisfy the Dirac equation.
A straightforward calculation gives
\begin{align}
i \gamma^{\mu} \partial_{\mu} \psi_{n}(x)
&= m \int\dfrac{d^{3}p}{(2\pi)^{3}}\frac{1}{\sqrt{2E_{\p}}}\sum_{\sigma}
\Bigl\{
b_{u, n} \Bigl[ e^{-ip\cdot x} u_{n}(\p,\sigma) a_{n}(\p,\sigma) \Bigr]
\nonumber \\
&\quad \quad \quad \quad \quad \quad \quad \quad \quad \quad \ 
- b_{v, n} \Bigl[ e^{ip\cdot x} v_{n}(\p,\sigma) a^{c \dag}_{n}(\p,\sigma) \Bigr]  \Bigr\}
\, ,
\label{eq:sf_W_21}
\end{align}
which implies that $\psi_{n}(x)$ is an eigenstate of the Dirac operator $i \gamma^{\mu} \partial_{\mu}$ if and only if 
\begin{align}
b_{n} \equiv b_{u, n} = - b_{v, n}
\, ,
\label{eq:sf_W_Dirac_con_0}
\end{align}
so that
\begin{align}
i \gamma^{\mu} \partial_{\mu} \psi_{n}(x)
&= m b_{n} \psi_{n}(x)
\, .
\label{eq:sf_W_Dirac_01}
\end{align}
In particular, the Dirac condition for $\psi_{n}(x)$ in Eq.~\eqref{eq:sf_W_Dirac_con_0} exactly coincides with the causality non-mixing condition in Eq.~\eqref{eq:causal_Dirac_b_1}.
Moreover, if the superposition field $\lambda(x)$~\eqref{eq:field_1+1_W_0} also satisfies the Dirac-like equation, the condition~\eqref{eq:sf_W_Dirac_con_0} must be extended to account for the Wigner degeneracy
\begin{align}
b \equiv b_{u,+\frac{1}{2}} = b_{u,-\frac{1}{2}} 
= -b_{v,+\frac{1}{2}} = -b_{v,-\frac{1}{2}} = \pm 1 
\, ,
\label{eq:sf_W_Dirac_con_s}
\end{align}
which explicitly violates the Elko condition~\eqref{eq:causal_elko_b_000}.
Then, taking the sum of Eq.~\eqref{eq:sf_W_21} over the Wigner degeneracy yields
\begin{align}
i \gamma^{\mu} \partial_{\mu} \lambda(x)
&= m b \, \lambda(x)
\, ,
\label{eq:sf_W_Dirac_s}
\end{align}
indicating that $\lambda(x)$ satisfies a Dirac-like equation with an overall sign factor $b$.
This factor carries over to the Hermitian-conjugate form of the equation for both the associated Dirac dual field $\bar{\lambda}(x)$~\eqref{eq:Dirac_dual_0} and the Elko dual field $\dual{\lambda}(x)$~\eqref{eq:elko_dual_0}
\begin{align}
-i \partial_{\mu} \bar{\lambda}(x) \gamma^{\mu} 
&= m b \, \bar{\lambda}(x)
\, ,
\label{eq:sf_W_Dirac_s_DD_0} 
\\
-i \partial_{\mu} \dual{\lambda}(x) \gamma^{\mu} 
&= m b \, \dual{\lambda}(x)
\, .
\label{eq:sf_W_Dirac_s_eD_0}
\end{align}
It is important to note that the Dirac-like equation for $\psi_{n}(x)$ is only valid under the Dirac condition~\eqref{eq:sf_W_Dirac_con_0}. However, irrespective of this condition, and using the on-shell relation $p^2=m^2$,
one can directly verify that
the Wigner-degenerate fields $\psi_{n}(x)$, $n=\pm \frac{1}{2}$~\eqref{eq:sf_W_20} always satisfy the Klein-Gordon equation: 
\begin{align}
\left( \partial^{\mu} \partial_{\mu} + m^2 \right) \psi_{n}(x)
= 0
\, .
\label{eq:sf_W_KG_0}
\end{align}
Thus, the superposition field $\lambda(x)$ also satisfies the Klein-Gordon equation
\begin{align}
\left( \partial^{\mu} \partial_{\mu} + m^2 \right) \lambda(x)
= 0
\, ,
\label{eq:sf_W_KG_s}
\end{align}
and this naturally extends to both the Dirac dual field $\bar{\lambda}(x)$~\eqref{eq:Dirac_dual_0} and the Elko dual field $\dual{\lambda}(x)$~\eqref{eq:elko_dual_0} 
\begin{align}
\left( \partial^{\mu} \partial_{\mu} + m^2 \right) \bar{\lambda}(x)
&= 0
\, ,
\label{eq:sf_W_KG_s_DD_0}
\\
\left( \partial^{\mu} \partial_{\mu} + m^2 \right) \dual{\lambda}(x)
&= 0
\, .
\label{eq:sf_W_KG_s_eD_0}
\end{align}

\subsection{Free Wigner-Dirac field}\label{sec:wigner_dirac_field}

In order to realize a Dirac-like Lagrangian density, we impose $b=1$ to the Dirac condition in Eq.~\eqref{eq:sf_W_Dirac_con_s}.
The Lorentz-invariant Dirac Lagrangian is therefore
\begin{align}
\mathcal{L}_{0} = \bar{\lambda} \left(i\gamma^{\mu} \partial_{\mu} - m  \right) \lambda 
\, ,
\label{eq:L_Dirac_free_D_0}
\end{align}
which generates the Dirac equations~\eqref{eq:sf_W_Dirac_s}-\eqref{eq:sf_W_Dirac_s_DD_0} with $b=1$ as the Euler-Lagrange equations for the free superposition field $\lambda(x)$~\eqref{eq:field_1+1_W_0}
and its Dirac dual fields $\bar{\lambda}(x)$~\eqref{eq:Dirac_dual_0} via the action principle. 
Setting $b=1$ also leads to the simplified ortho-normalization relations from Eqs.~\eqref{eq:norm_Dirac_u_1}-\eqref{eq:norm_Dirac_uv_m_3}:
\begin{align}
\bar{u}(\p,\sigma) u(\p,\sigma') &= -\bar{v}(\p,\sigma) v(\p,\sigma') = 2m \delta_{\sigma\sigma'} \, ,
\label{eq:norm_uv_D_1} \\
\bar{u}(\p,\sigma) v(\p,\sigma') &= u^{\dag}(\p,\sigma) v(-\p,\sigma')
= \bar{v}(\p,\sigma) u(\p,\sigma') = v^{\dag}(\p,\sigma) u(-\p,\sigma') = 0 \, ,
\label{eq:norm_uv_m_D_1}
\end{align}
with the universal notations $u(\p,\sigma) \equiv u_{+\frac{1}{2}}(\p,\sigma)= u_{-\frac{1}{2}}(\p,\sigma)$ and $v(\p,\sigma) \equiv v_{+\frac{1}{2}}(\p,\sigma)= v_{-\frac{1}{2}}(\p,\sigma)$.
The canonical momentum field conjugate to $\lambda(x)$~\eqref{eq:field_1+1_W_0} is defined as
\begin{align}
\pi(x) \equiv \dfrac{\partial \mathcal{L}_{0}}{\partial \dot{\lambda}(x)} 
= i\lambda^{\dag}(x)
\, , 
\label{eq:Dirac_pi_free_D_0}
\end{align}
which can be written explicitly in the Fourier modes
\begin{align}
\pi (x) 
= \int\dfrac{d^{3}p}{(2\pi)^{3}}\frac{i}{2\sqrt{E_{\p}}}\sum_{n,\sigma} \left[ e^{ip\cdot x} u^{\dag}(\p,\sigma)a_{n}^{\dag}(\p,\sigma)
    + e^{-ip\cdot x} v^{\dag}(\p,\sigma)a^{c}_{n}(\p,\sigma) \right]  
\label{eq:Dirac_pi_free_D_1}
\, .
\end{align}
Using the ortho-normalization relations~\eqref{eq:norm_uv_D_1}-\eqref{eq:norm_uv_m_D_1}, the canonical equal-time anticommutation relations follow immediately 
\begin{align}
\bigl\{\lambda_{\ell}(t, \x), \pi_{\ell'}(t, \x')  \bigr\} &= i\delta^{(3)}(\x - \x') \delta_{\ell \ell'} \, ,
\label{eq:can_W_field_1} \\
\bigl\{\lambda_{\ell}(t, \x), \lambda_{\ell'}(t, \x')  \bigr\} &= 
\bigl\{\lambda^{\dag}_{\ell}(t, \x), \lambda^{\dag}_{\ell'}(t, \x')  \bigr\} = 0 \, ,
\label{eq:can_Dirac_field_D_0}
\end{align}
which are identical to those for the standard free Dirac spinor fields, since we have the standard polarizations and anticommuting relations for different Wigner degeneracies~\eqref{eq:can_W_ab_1}-\eqref{eq:can_W_ab_2}. 
After performing the Legendre transformation, it is straightforward to show that the Lagrangian density of Eq.~\eqref{eq:L_Dirac_free_D_0}
yields the Hamiltonian 
\begin{align}
    H_{0} &= \int d^3x \left[ \pi(x) \dot{\lambda}(x) - \mathcal{L}_{0}(x) \right]
    \nonumber \\
    &= \int d^3x \ \bar{\lambda}(x)(-i\gamma^{i}\partial_{i}+m) \lambda(x)
    \nonumber \\
    &= \dfrac{1}{2} \sum_{n} \int d^3x \ \bar{\psi}_{n}(x)(-i\gamma^{i}\partial_{i}+m) \Bigl[\psi_{n}(x) + \psi_{-n}(x) \Bigr]
    \, ,
\label{eq:H_Dirac_free_D_0}
\end{align}
which contains an additional term mixing the Wigner degeneracies compared to the standard Hamiltonian in the conventional QFT
\begin{align}
\dfrac{1}{2} \sum_{n} \int d^3x \ \bar{\psi}_{n}(x)(-i\gamma^{i}\partial_{i}+m) \psi_{-n}(x)
=
\dfrac{i}{2} \sum_{n} \int d^3x \ \psi^{\dag}_{n}(x) \dot{\psi}_{-n}(x)
    \, ,
\label{eq:H_Dirac_free_D_1}
\end{align}
where the Dirac equation for $\psi_{-n}(x)$ has been inserted.
Inserting quantization of $\psi_{n}(x)$~\eqref{eq:sf_W_20} into Eq.~\eqref{eq:H_Dirac_free_D_1}, we find
\begin{align}
\dfrac{i}{2} \int d^3x \ \psi^{\dag}_{n}(x) \dot{\psi}_{-n}(x) 
= \int \dfrac{d^{3}p}{(2\pi)^{3}} 
\dfrac{E_{\p}}{2}
\sum_{\sigma }
\Bigl[
a_{n}^{\dag}(\p,\sigma) a_{-n}(\p,\sigma)
-
a^{c}_{n}(\p,\sigma) a^{c \dag}_{-n}(\p,\sigma)
\Bigr]
\, ,
\label{eq:H_Dirac_free_D_2}
\end{align}
confirming a wrong mixing between different Wigner degeneracies. 
Thus, this naive construction with the free superposition field $\lambda(x)$
and its Dirac dual field $\bar{\lambda}(x)$ is not correct. 
To address this issue, we need to introduce a Wigner doublet field
\begin{equation}
    \Psi(x)\equiv\left[\begin{matrix}
        \psi_{+\frac{1}{2}}(x) \\
        \psi_{-\frac{1}{2}}(x)
    \end{matrix}\right], 
    \label{eq:doublet}
\end{equation}
along with the free Lagrangian density 
\begin{equation}
    \mathcal{L}_{0} = \overline{\Psi} (i\gamma^{\mu}\partial_{\mu}-m) \Psi 
    \, .
    \label{eq:LPsi_1}
\end{equation}
Performing the Legendre transformation yields the Hamiltonian
\begin{align}
    H_{0} = \int d^3x \ \overline{\Psi}(x)(-i\gamma^{i}\partial_{i}+m)\Psi(x)
    \, .
\label{eq:HPsi_1}
\end{align}
Upon Fourier expanding the two Wigner-degenerate fields $\psi_{n}(x)$, $n=\pm\frac{1}{2}$~\eqref{eq:sf_W_20}, and applying normal ordering, we obtain the correct Hamiltonian as expected~\cite{deGracia:2024yei}
\begin{align}
    H_{0} &= \int\frac{d^{3}p}{(2\pi)^{3}} \sum_{n,\sigma} E_{\p} \left[a^{\dag}_{n}(\p,\sigma) a_{n}(\p,\sigma)+a^{c\dag}_{n}(\p,\sigma) a^{c}_{n}(\p,\sigma)\right]
    \, ,
\label{eq:HPsi_2}
\end{align}
A detailed analysis of this construction is provided in our previous work~\cite{Lee:2025kpz}.

\subsection{Free Wigner-Klein-Gordon field}
\label{sec:W_KG_1}

In this section, we examine the free Klein-Gordon Lagrangian density for the free Wigner superposition field $\lambda(x)$~\eqref{eq:field_1+1_W_0}
and its Dirac dual field $\bar{\lambda}(x)$~\eqref{eq:Dirac_dual_0}:
\begin{align}
\mathcal{L}_{0} = \partial^{\mu} \bar{\lambda} \partial_{\mu} \lambda - m^2 \bar{\lambda} \lambda 
\, .
\label{eq:L_Dirac_free_KG_0}
\end{align}
For convenience, the canonical field $\lambda (x)$ is rescaled by the factor $\sqrt{m}$ 
\begin{align}
\lambda (x) \longrightarrow \dfrac{\lambda (x)}{\sqrt{m}}
\, ,
\label{eq:Dirac_rescale_0}
\end{align}
so that its mass dimension is shifted to $[\lambda]=1$.
Rather than specifying a particular configuration for $b_{u, n}$ and $b_{v, n}$ satisfying the causality condition~\eqref{eq:causal_Dirac_b_0}, we leave them general for now. 
The associated canonical conjugate momentum fields are defined as
\begin{align}
\pi (x) \equiv \dfrac{\partial \mathcal{L}_{0}}{\partial \dot{\lambda} (x)} 
= \dot{\bar{\lambda}} (x)
\, , \quad 
\bar{\pi} (x) \equiv \dfrac{\partial \mathcal{L}_{0}}{\partial \dot{\bar{\lambda}} (x)} 
= -\dot{\lambda} (x)
\, ,
\label{eq:Dirac_pi_free_KG_0}
\end{align}
which can be quantized explicitly with Eqs.~\eqref{eq:field_1+1_W_0} and \eqref{eq:Dirac_dual_1}
\begin{align}
\pi (x) 
&= \frac{i}{2} \int\dfrac{d^{3}p}{(2\pi)^{3}}\sqrt{\frac{E_{\p}}{m}}\sum_{n,\sigma} \left[ e^{ip\cdot x} \bar{u}_{n}(\p,\sigma)a_{n}^{\dag}(\p,\sigma)
    - e^{-ip\cdot x} \bar{v}_{n}(\p,\sigma)a^{c}_{n}(\p,\sigma) \right] 
\, , 
\label{eq:Dirac_pi_free_KG_11} \\
\bar{\pi} (x) 
&= \frac{i}{2} \int\dfrac{d^{3}p}{(2\pi)^{3}}\sqrt{\frac{E_{\p}}{m}}\sum_{n,\sigma}\left[e^{-ip\cdot x}u_{n}(\p,\sigma)a_{n}(\p,\sigma)-e^{ip\cdot x} v_{n}(\p,\sigma)a^{c \dag}_{n}(\p,\sigma)\right] 
\, ,
\label{eq:Dirac_pi_free_KG_12}
\end{align}
where we have inserted the rescale factor $\sqrt{m}$ as indicated in Eq.~\eqref{eq:Dirac_rescale_0}.
The canonical equal-time anticommutation relations~\footnote{We rescale the field $\lambda (x)$ with the factor $\sqrt{m}$ following Eq.~\eqref{eq:Dirac_rescale_0} by default.} are calculated straightforward from anticommutation relations of creation and annihilation operators~\eqref{eq:can_W_ab_1}-\eqref{eq:can_W_ab_2}, and sums of polarizations~\eqref{eq:sum_Dirac_u_0}-\eqref{eq:sum_Dirac_v_0}
\begin{align}
\left\{ \lambda_{\ell} (t, \x), \pi_{\ell'} (t, \y) \right\}
&= \frac{i}{m} \int\dfrac{d^{3}p}{(2\pi)^{3}} e^{-i \p \cdot (\x - \y)}
\left(\p \cdot \boldsymbol{\gamma}_{\ell \ell'} \right) 
+
\frac{i}{2} b_{u} \,
\delta^{(3)}(\x-\y) \delta_{\ell \ell'}
\nonumber \\
&\neq i \delta^{(3)}(\x-\y) \delta_{\ell \ell'}
\, ,  
\label{eq:can_Dirac_KG_1} \\
\bigl\{ \bar{\lambda}_{\ell} (t, \x), \bar{\pi}_{\ell'} (t, \y) \bigl\}
&= \frac{i}{m} \int\dfrac{d^{3}p}{(2\pi)^{3}} e^{i \p \cdot (\x - \y)}
\left(\p \cdot \boldsymbol{\gamma}_{\ell' \ell} \right) 
+
\frac{i}{2} b_{u} \,
\delta^{(3)}(\x-\y) \delta_{\ell \ell'}
\nonumber \\
&\neq i \delta^{(3)}(\x-\y) \delta_{\ell \ell'}
\, ,  
\label{eq:can_Dirac_KG_2} \\
\bigl\{ \lambda_{\ell} (t, \x), \bar{\lambda}_{\ell'} (t, \y) \bigl\}
&= \int\dfrac{d^{3}p}{(2\pi)^{3}} e^{i \p \cdot (\x - \y)}
\gamma^{0}_{\ell \ell'} 
\neq 0
\, ,  
\label{eq:can_Dirac_KG_3} \\ 
\bigl\{ \pi_{\ell} (t, \x), \bar{\pi}_{\ell'} (t, \y) \bigl\}
&= -\frac{1}{m} \int\dfrac{d^{3}p}{(2\pi)^{3}} e^{i \p \cdot (\x - \y)} E_{\p}^2
\gamma^{0}_{\ell \ell'} 
\neq 0
\, ,  
\label{eq:can_Dirac_KG_4}  
\end{align}
and
\begin{align}
\bigl\{ \lambda_{\ell} (t, \x), \lambda_{\ell'} (t, \y) \bigl\}
&= \bigl\{ \bar{\lambda}_{\ell} (t, \x), \bar{\lambda}_{\ell'} (t, \y) \bigl\}
= 0
\, ,
\label{eq:can_Dirac_KG_5} \\
\bigl\{ \pi_{\ell} (t, \x), \pi_{\ell'} (t, \y) \bigl\}
&= \bigl\{ \bar{\pi}_{\ell} (t, \x), \bar{\pi}_{\ell'} (t, \y) \bigl\}
= 0
\, ,
\label{eq:can_Dirac_KG_6} \\
\bigl\{ \lambda_{\ell} (t, \x), \bar{\pi}_{\ell'} (t, \y) \bigl\}
&= \bigl\{ \bar{\lambda}_{\ell} (t, \x), \pi_{\ell'} (t, \y) \bigl\}
= 0
\, .
\label{eq:can_Dirac_KG_7} 
\end{align}
These results show that the canonical quantization can not be realized by adjusting factors $b_{u, n}$ and $b_{v, n}$ in
the Klein-Gordon framework for $\lambda(x)$~\eqref{eq:field_1+1_W_0} and $\bar{\lambda}(x)$~\eqref{eq:Dirac_dual_0}.
Thus, this approach is untenable and must be eliminated in favor of alternative constructions that can consistently accommodate the canonical formalism.

\subsection{Free Elko-Dirac field}

We now consider the situation where the kinematics of $\lambda(x)$~\eqref{eq:field_1+1_W_0} and its Elko dual field $\dual{\lambda}(x)$~\eqref{eq:elko_dual_0} are described by the Dirac Lagrangian density
\begin{align}
\mathcal{L}_{0}=\dual{\lambda}(i\gamma^{\mu}\partial_{\mu}-m)\lambda\,,\label{eq:L_Elko_Dirac}
\end{align}
which implies that both $\lambda(x)$ and $\dual{\lambda}(x)$ satisfy the Dirac equation~\eqref{eq:sf_W_Dirac_s} with $b=1$ as their equations of motion. The canonical momentum conjugate to $\lambda (x)$ is given by
\begin{equation}
\pi(x) \equiv \frac{\partial\mathcal{L}_{0}}{\partial\dot{\lambda}(x)}=i\dual{\lambda}(x)\gamma^{0}
= i\lambda^{\dag}(x) \, , 
\label{eq:elko_pi_free_Dirac_0}
\end{equation}
where the last equality is derived by inserting the Dirac equation for $\lambda (x)$ to the relationship between $\dual{\lambda}(x)$ and $\bar{\lambda}(x)$ given in Eq.~\eqref{eq:elko_dual_0}, i.e., $\dual{\lambda}(x) = \bar{\lambda}(x)$. 
Thus, the canonical equal-time anticommutation relations associated to $\mathcal{L}_{0}$~\eqref{eq:L_Elko_Dirac} with the Elko dual field $\dual{\lambda}(x)$~\eqref{eq:elko_dual_0} are identical to those in Eqs.~\eqref{eq:can_W_field_1}-\eqref{eq:can_Dirac_field_D_0} induced by the Dirac dual field $\bar{\lambda}(x)$~\eqref{eq:Dirac_dual_0}.
Similarly, performing the Legendre transformation on the Elko-Dirac $\mathcal{L}_{0}$~\eqref{eq:L_Elko_Dirac} leads to a Hamiltonian that matches Eq.~\eqref{eq:H_Dirac_free_D_0}. Although the Elko-Dirac Lagrangian density correctly reproduces the canonical anticommutation relations, it yields an incorrect free Hamiltonian~\eqref{eq:H_Dirac_free_D_0}. 
Therefore, the Elko-Dirac construction must also be ruled out as a viable formulation.


\subsection{Free Elko-Klein-Gordon field}
\label{sec:elko_KG_1}

In this section, we construct the free Klein-Gordon Lagrangian density for the Wigner superposition field $\lambda(x)$~\eqref{eq:field_1+1_W_0} and its Elko dual fields $\dual{\lambda}(x)$~\eqref{eq:elko_dual_0}: 
\begin{align}
\mathcal{L}_{0} = \partial^{\mu} \dual{\lambda} \partial_{\mu} \lambda - m^2 \dual{\lambda} \lambda 
\, ,
\label{eq:L_elko_free_KG_0}
\end{align}
which yields the Euler-Lagrange equations for canonical fields $\lambda (x)$ and $\dual{\lambda} (x)$, confirming that they satisfy the Klein-Gordon equation, regardless of the specific choices of the sign factors $b_{u, n}$ and $b_{v, n}$ in Eqs.~\eqref{eq:sf_W_KG_s} and \eqref{eq:sf_W_KG_s_eD_0}. 
As in the previous Section~\ref{sec:W_KG_1}, for convenience in the Klein-Gordon framework, we rescale the canonical field $\lambda (x)$ by the factor $\sqrt{m}$ 
\begin{align}
\lambda (x) \longrightarrow \dfrac{\lambda (x)}{\sqrt{m}}
\, ,
\label{eq:elko_rescale_0}
\end{align}
such that its mass dimension becomes $[\lambda]=1$.
The corresponding canonical conjugate momentum fields are defined as~\footnote{Note that $\dual{\pi} (x)$ is not equal to $\pi (x)$ after applying the Elko dual operation~\eqref{eq:elko_dual_0}. The minus sign of $\dual{\pi} (x)$ comes from anticommuting fields $\lambda (x)$ and $\dual{\lambda} (x)$.}
\begin{align}
\pi (x) \equiv \dfrac{\partial \mathcal{L}_{0}}{\partial \dot{\lambda} (x)} 
= \dot{\dual{\lambda}} (x)
\, , \quad 
\dual{\pi} (x) \equiv \dfrac{\partial \mathcal{L}_{0}}{\partial \dot{\dual{\lambda}} (x)} 
= -\dot{\lambda} (x)
\, ,
\label{eq:elko_pi_free_KG_0}
\end{align}
which can be decomposed into Fourier modes explicitly via Eqs.~\eqref{eq:field_1+1_W_0} and \eqref{eq:elko_dual_1}
\begin{align}
\pi (x) 
&= \frac{i}{2} \int\dfrac{d^{3}p}{(2\pi)^{3}}\sqrt{\frac{E_{\p}}{m}}\sum_{n,\sigma} \left[ e^{ip\cdot x} \dual{u}_{n}(\p,\sigma)a_{n}^{\dag}(\p,\sigma)
    - e^{-ip\cdot x} \dual{v}_{n}(\p,\sigma)a^{c}_{n}(\p,\sigma) \right] 
\, , 
\label{eq:elko_pi_free_KG_11} \\
\dual{\pi} (x) 
&= \frac{i}{2} \int\dfrac{d^{3}p}{(2\pi)^{3}}\sqrt{\frac{E_{\p}}{m}}\sum_{n,\sigma}\left[e^{-ip\cdot x}u_{n}(\p,\sigma)a_{n}(\p,\sigma)-e^{ip\cdot x} v_{n}(\p,\sigma)a^{c \dag}_{n}(\p,\sigma)\right] 
\, ,
\label{eq:elko_pi_free_KG_12}
\end{align}
where we have inserted the rescale factor $\sqrt{m}$ as indicated in Eq.~\eqref{eq:elko_rescale_0}.
The canonical equal-time anticommutation relations~\footnote{We rescale the field $\lambda (x)$ with the factor following $\sqrt{m}$~\eqref{eq:elko_rescale_0} by default.} are calculated straightforward from anticommutation relations of creation and annihilation operators~\eqref{eq:can_W_ab_1}-\eqref{eq:can_W_ab_2}, and sums of polarizations~\eqref{eq:sum_elko_u_12}-\eqref{eq:sum_elko_v_12}
\begin{align}
\bigl\{ \lambda_{\ell} (t, \x), \pi_{\ell'} (t, \y) \bigr\}
&= \frac{i b_{u}}{2m} \int\dfrac{d^{3}p}{(2\pi)^{3}} \ 
e^{-i \p \cdot (\x - \y)} \left(\p \cdot \boldsymbol{\gamma}_{\ell \ell'} \right) 
+ i \delta^{(3)}(\x-\y) \delta_{\ell \ell'}
\, ,  
\label{eq:can_elko_1} \\
\bigl\{ \dual{\lambda}_{\ell} (t, \x), \dual{\pi}_{\ell'} (t, \y) \bigr\}
&= \frac{i b_{u}}{2m} \int\dfrac{d^{3}p}{(2\pi)^{3}} \ 
e^{i \p \cdot (\x - \y)} \left(\p \cdot \boldsymbol{\gamma}_{\ell' \ell} \right) 
+ i \delta^{(3)}(\x-\y) \delta_{\ell \ell'}
\, ,  
\label{eq:can_elko_2} \\ 
\bigl\{ \lambda_{\ell} (t, \x), \dual{\lambda}_{\ell'} (t, \y) \bigr\}
&= \frac{b_{u}}{2} \int\dfrac{d^{3}p}{(2\pi)^{3}} e^{i \p \cdot (\x - \y)}
\gamma^{0}_{\ell \ell'} 
\, ,  
\label{eq:can_elko_3}  \\
\bigl\{ \pi_{\ell} (t, \x), \dual{\pi}_{\ell'} (t, \y) \bigr\}
&= -\frac{b_{u}}{2m} \int\dfrac{d^{3}p}{(2\pi)^{3}} e^{i \p \cdot (\x - \y)} E_{\p}^2
\gamma^{0}_{\ell \ell'} 
\, ,  
\label{eq:can_elko_4}  
\end{align}
and
\begin{align}
\bigl\{ \lambda_{\ell} (t, \x), \lambda_{\ell'} (t, \y) \bigr\}
&= \bigl\{ \dual{\lambda}_{\ell} (t, \x), \dual{\lambda}_{\ell'} (t, \y) \bigr\}
= 0
\, ,
\label{eq:can_elko_5} \\
\bigl\{ \pi_{\ell} (t, \x), \pi_{\ell'} (t, \y) \bigr\}
&= \bigl\{ \dual{\pi}_{\ell} (t, \x), \dual{\pi}_{\ell'} (t, \y) \bigr\}
= 0
\, ,
\label{eq:can_elko_6} \\
\bigl\{ \lambda_{\ell} (t, \x), \dual{\pi}_{\ell'} (t, \y) \bigr\}
&= \bigl\{ \dual{\lambda}_{\ell} (t, \x), \pi_{\ell'} (t, \y) \bigr\}
= 0
\, .
\label{eq:can_elko_7} 
\end{align}
These results demonstrate that the canonical quantization can be successfully implemented within the Klein-Gordon framework for the fields $\lambda(x)$ and $\dual{\lambda}(x)$ if and only if $b_{u}=b_{v}=0$ (or equivalently $\Delta = -1$), which coincides with the Elko condition~\eqref{eq:causal_elko_b_000}-\eqref{eq:causal_elko_b_001}, but explicitly violates the Dirac condition for $\lambda(x)$ given in Eq.~\eqref{eq:sf_W_Dirac_con_s}~\footnote{Note that the Dirac condition for the Wigner-degenerate field $\psi_{n}(x)$ in Eq.~\eqref{eq:sf_W_Dirac_con_0} may still be fulfilled.}.
The Hamiltonian for the free Elko fields --- the Wigner superposition field under the Elko condition --- is derived via the Legendre
transformation from the free Lagrangian $\mathcal{L}_{0}$~\eqref{eq:L_elko_free_KG_0}
\begin{align}
H_0 &= \int d^3x \left[ \pi \dot{\lambda} + \dual{\pi} \dot{\dual{\lambda}} - \mathcal{L}_{0} \right] 
\nonumber \\
&= \int d^3x \left[ \dot{\dual{\lambda}} \dot{\lambda} - \dot{\lambda} \dot{\dual{\lambda}} - \partial^{\mu} \dual{\lambda} \partial_{\mu} \lambda + m^2 \dual{\lambda} \lambda  \right] 
\nonumber \\
&= \int d^3x \left[ \dot{\dual{\lambda}} \dot{\lambda} + \nabla \dual{\lambda} \cdot \nabla \lambda + m^2 \dual{\lambda} \lambda  \right]
\, ,
\label{eq:H_elko_free_0}
\end{align}
where we have imposed the conjugate momenta~\eqref{eq:elko_pi_free_KG_0} and an extra sign in the third step arises due to the anticommuting nature of the fermionic fields $\dot{\dual{\lambda}}$ and $\dot{\lambda}$.
We are now ready to write $H_0$ in terms of ladder operators~\eqref{eq:can_W_ab_1}-\eqref{eq:can_W_ab_2}.
Inserting the rescaled fields $\lambda (x)$~\eqref{eq:field_1+1_W_0} and $\dual{\lambda} (x)$~\eqref{eq:elko_dual_1} to Eq.~\eqref{eq:H_elko_free_0}, we find~\cite{deGracia:2024yei}
\begin{align}
    H_{0} &= \int\frac{d^{3}p}{(2\pi)^{3}} \sum_{n,\sigma} E_{\p} \left[a^{\dag}_{n}(\p,\sigma) a_{n}(\p,\sigma)+a^{c\dag}_{n}(\p,\sigma) a^{c}_{n}(\p,\sigma)\right]
    \, ,
\label{eq:H_elko_free_1}
\end{align}
where we have omitted an infinite zero-point energy shift and imposed
\begin{align}
\int d^3x \ \dot{\dual{\lambda}} \dot{\lambda} 
=& \frac{1}{2} \int \dfrac{d^{3}p}{(2\pi)^{3}} E_{\p} 
\sum_{n,\sigma}
\Bigl[
a_{n}^{\dag}(\p,\sigma) a_{n}(\p,\sigma)
-
a^{c}_{n}(\p,\sigma) a^{c \dag}_{n}(\p,\sigma)
\Bigr]
\nonumber \\
&-\frac{1}{4m} \int \dfrac{d^{3}p}{(2\pi)^{3}}  E_{\p} 
\sum_{n n',\sigma \sigma'}
\Bigl[
e^{2iE_{\p} t} 
\dual{u}_{n}(\p,\sigma) v_{n'}(-\p,\sigma')
a_{n}^{\dag}(\p,\sigma) a^{c \dag}_{n'}(-\p,\sigma')
\nonumber \\
&\quad \quad \quad \quad \quad \ 
+e^{-2iE_{\p} t} 
\dual{v}_{n}(\p,\sigma) u_{n'}(-\p,\sigma')
a^{c}_{n}(\p,\sigma) a_{n'}(-\p,\sigma')
\Bigr]
\, , \\
\int d^3x \ m^2 \dual{\lambda} \lambda
=& \frac{1}{2} \int \dfrac{d^{3}p}{(2\pi)^{3}} 
\frac{m^2}{E_{\p}}
\sum_{n,\sigma}
\Bigl[
a_{n}^{\dag}(\p,\sigma) a_{n}(\p,\sigma)
-
a^{c}_{n}(\p,\sigma) a^{c \dag}_{n}(\p,\sigma)
\Bigr]
\nonumber \\
&+\frac{1}{4m} \int \dfrac{d^{3}p}{(2\pi)^{3}}  
\frac{m^2}{E_{\p}}
\sum_{n n',\sigma \sigma'}
\Bigl[
e^{2iE_{\p} t} 
\dual{u}_{n}(\p,\sigma) v_{n'}(-\p,\sigma')
a_{n}^{\dag}(\p,\sigma) a^{c \dag}_{n'}(-\p,\sigma')
\nonumber \\
&\quad \quad \quad \quad \quad \ 
+e^{-2iE_{\p} t} 
\dual{v}_{n}(\p,\sigma) u_{n'}(-\p,\sigma')
a^{c}_{n}(\p,\sigma) a_{n'}(-\p,\sigma')
\Bigr]
\, , \\
\int d^3x \ \nabla \dual{\lambda} \cdot \nabla \lambda
=& \frac{1}{2} \int \dfrac{d^{3}p}{(2\pi)^{3}} 
\frac{\p^2}{E_{\p}}
\sum_{n,\sigma}
\Bigl[
a_{n}^{\dag}(\p,\sigma) a_{n}(\p,\sigma)
-
a^{c}_{n}(\p,\sigma) a^{c \dag}_{n}(\p,\sigma)
\Bigr]
\nonumber \\
&+\frac{1}{4m} \int \dfrac{d^{3}p}{(2\pi)^{3}}  
\frac{\p^2}{E_{\p}}
\sum_{n n',\sigma \sigma'}
\Bigl[
e^{2iE_{\p} t} 
\dual{u}_{n}(\p,\sigma) v_{n'}(-\p,\sigma')
a_{n}^{\dag}(\p,\sigma) a^{c \dag}_{n'}(-\p,\sigma')
\nonumber \\
&\quad \quad \quad \quad \quad  \     
+e^{-2iE_{\p} t} 
\dual{v}_{n}(\p,\sigma) u_{n'}(-\p,\sigma')
a^{c}_{n}(\p,\sigma) a_{n'}(-\p,\sigma')
\Bigr]
\, ,
\end{align}
using the ortho-normalization relations~\eqref{eq:norm_elko_uv_2}. As expected, $H_0$ reduces to a sum over all Wigner-degenerate particle and antiparticle modes. This result validates the canonical quantization procedure for the Elko field and reflects its underlying Klein-Gordon structure.
Furthermore, by imposing Eqs.~\eqref{eq:S_re_T_field_s2} and \eqref{eq:T_W_field_s_elko_s0} (with $\Delta = -1$) to the free Klein-Gordon Lagrangian density $\mathcal{L}_{0}$~\eqref{eq:L_elko_free_KG_0}, we obtain
\begin{align}
\hat{S}_{T} \mathcal{L}_{0}(x) \hat{S}_{T}^{-1} =\mathcal{L}_{0} (x) \, ,
\label{eq:S_re_T_elko_Lag_0_2} 
\end{align}
implying that a free Klein-Gordon theory for the Elko field is always invariant under the internal $\hat{S}_{T}$ symmetry.
Therefore, for a complete theory that fully incorporates the physical two-fold Wigner degeneracy, this internal $\hat{S}_{T}$ symmetry must be broken in the interacting sector via a mechanism yet to be identified. This requirement aligns with the doublet framework proposed in Ref.~\cite{Lee:2025kpz}, and determining such a mechanism will be one of the central goals in our future investigations into Elko interactions.
Additionally, the intrinsic mismatch in mass dimension between the Elko field and the SM matter fields naturally suppresses their interactions. This suppression provides a fundamental origin for the ``darkness'' of Elko, making it a viable DM candidate~\cite{Ahluwalia:2022ttu}. Crucially, this feature is not imposed ad hoc but emerges directly from the structure of the Elko theory, offering a compelling and natural explanation for the elusive nature of Elko interactions.


To facilitate comparison with commonly used forms in the literature, let us first consider the initial basis with the associated polarizations~\eqref{eq:+u}-\eqref{eq:-v} in the non-zero causality mixing configuration given by Eq.~\eqref{eq:causal_Dirac_b_2} with $+1$, which explicitly violates the Dirac conditions~\eqref{eq:sf_W_Dirac_con_0}-\eqref{eq:sf_W_Dirac_con_s}~\footnote{An equivalent analysis can be performed for the case with $-1$. However, to achieve the same transformed polarizations, we need to adapt the transformation matrix $S$ in Eq.~\eqref{eq:elko_spin2C_0}.}.
To derive a practical representation commonly used in the literature, we impose the following specific transformation matrix to Eq.~\eqref{eq:deg_op_ab2C_1} 
\begin{align}
S=S^{-1} =
\frac{1}{2}
\left[\begin{matrix}
\mathds{1}+\sigma^2 & \mathds{1}-\sigma^2 \\
\mathds{1}-\sigma^2 & -\mathds{1}-\sigma^2
\end{matrix}\right] = S^{\dag} = S^{c}
\, .
\label{eq:elko_spin2C_0} 
\end{align}
In the rest frame, the explicit form of the new polarizations $\xi_{\tau}(\0)$ and $\zeta_{\tau}(\0)$, with $\tau=1,\cdots, 4$ in the new basis generated by the transformation in Eq.~\eqref{eq:elko_spin2C_0} can be derived by inserting Eqs.~\eqref{eq:+u}-\eqref{eq:-v} to Eqs.~\eqref{eq:elko_xi_0}-\eqref{eq:elko_zeta_0}:
\begin{alignat}{4}
\xi_{1}(\0) &= 
\sqrt{m}\left[\begin{matrix}
    1 \\
    0 \\
    0 \\
    i
\end{matrix}\right]
, \, \  
&\xi_{2}(\0) &=
\sqrt{m}\left[\begin{matrix}
    0 \\
    1 \\
    -i \\
    0
\end{matrix}\right] , \, \ 
&\xi_{3}(\0) &= 
\sqrt{m}\left[\begin{matrix}
    0 \\
    -i \\
    1 \\
    0
\end{matrix}\right] , \, \   
&\xi_{4}(\0) &= 
\sqrt{m}\left[\begin{matrix}
    i \\
    0 \\
    0 \\
    1
\end{matrix}\right] , 
\label{eq:elko_C_xi} 
\\
\zeta_{1}(\0) &= 
\sqrt{m}\left[\begin{matrix}
    0 \\
    1 \\
    i \\
    0
\end{matrix}\right]
\, , \quad  
&\zeta_{2}(\0) &= 
\sqrt{m}\left[\begin{matrix}
    -1 \\
    0 \\
    0 \\
    i
\end{matrix}\right] , \, \  
&\zeta_{3}(\0) &= 
\sqrt{m}\left[\begin{matrix}
    -i \\
    0 \\
    0 \\
    1
\end{matrix}\right] , \, \  
&\zeta_{4}(\0) &= 
\sqrt{m}\left[\begin{matrix}
    0 \\
    -i \\
    -1 \\
    0
\end{matrix}\right] . 
\label{eq:elko_C_zeta}
\end{alignat} 
The spinors $\xi_{\tau}(\p)$ and $\zeta_{\tau}(\p)$ have two notable features. First, their norms vanish under the Dirac dual, i.e., $\bar{\xi}_{\tau}(\p)\xi_{\tau}(\p)=0$, $\bar{\zeta}_{\tau}(\p)\zeta_{\tau}(\p)=0$. Second, in the rest frame $\p=\mathbf{0}$, their left- and right-handed components have opposite eigenvalues with respect to the spin-projection operator $\frac{1}{2}\sigma_{z}$. It was traditionally believed that these are the two features that distinguish Elko from the Dirac spinors~\cite{Ahluwalia:2023slc}. However, as we have demonstrated in this work, both features are artifacts of the choice of basis. As shown in Section~\ref{wigner_spinor_field}, the spinors $u_{n}(\p,\sigma)$ and $v_{n}(\p,\sigma)$, which are used to construct mass dimension one fields, have non-vanishing norms under the Dirac dual~\eqref{eq:norm_Dirac_u_1}-\eqref{eq:norm_Dirac_v_1}. Moreover, in the rest frame $\p=\mathbf{0}$, their left- and right-handed components have the same eigenvalues with respect to $\frac{1}{2}\sigma_{z}$.
On the other hand, for the initial basis with the associated polarizations~\eqref{eq:+u}-\eqref{eq:-v} in the configuration~\eqref{eq:causal_Dirac_b_1} without causality mixing, the same new basis can be achieved by first transforming it into the causality mixing configuration via the basis transformation in Eq.~\eqref{eq:cau_m2n_0}, as illustrated at the end of Section~\ref{sec:basisT_0}. The basis transformation $S$ in Eq.~\eqref{eq:elko_spin2C_0} can then be applied subsequently. This procedure can be formulated explicitly using the following basis transformation matrix for anti-particles~\footnote{Here we set $b_{u, +\frac{1}{2}} = -b_{v, +\frac{1}{2}} = -b_{u, -\frac{1}{2}} = b_{v, -\frac{1}{2}} = +1$.}:
\begin{align}
S^{c} &= \gamma^{0} S 
=
\frac{1}{2}
\left[\begin{matrix}
0 \quad  & \quad \mathds{1}  \\
\mathds{1} \quad  & \quad 0 
\end{matrix}\right]
\left[\begin{matrix}
\mathds{1}+\sigma^2 & \mathds{1}-\sigma^2 \\
\mathds{1}-\sigma^2 & -\mathds{1}-\sigma^2
\end{matrix}\right]
=
\frac{1}{2}
\left[\begin{matrix}
\mathds{1}-\sigma^2 & -\mathds{1}-\sigma^2 \\
\mathds{1}+\sigma^2 & \mathds{1}-\sigma^2
\end{matrix}\right]
\, ,
\label{eq:elko_spin2C_c1} 
\end{align}
so that
\begin{align}
S^{c-1} &= S^{c \dag} 
= S \gamma^{0}
=
\frac{1}{2}
\left[\begin{matrix}
\mathds{1}-\sigma^2 & \mathds{1}+\sigma^2 \\
-\mathds{1}-\sigma^2 & \mathds{1}-\sigma^2
\end{matrix}\right]
\, ,
\label{eq:elko_spin2C_c2} 
\end{align}
while the transformation matrix $S$ in Eq.~\eqref{eq:elko_spin2C_0} remains applied to particles.

Finally, we conclude that this new basis introduced above precisely reproduces the one originally defined via charge conjugation in Ref.~\cite{Ahluwalia:2023slc}~\footnote{Note that our chirality convention differs from that used in Ref.~\cite{Ahluwalia:2023slc}.}. This also clarifies the reason why the Wigner superposition field $\lambda (x)$~\eqref{eq:field_1+1_W_0} under the Elko condition~\eqref{eq:causal_elko_b_000} is referred to as the Elko field. 
In this work, the Elko field arises naturally through construction in terms of creation and annihilation operators corresponding to eigenstates of both the Wigner degeneracy and spin --- structures that generate directly from Lorentz covariance.

\section{Summary and conclusions}

Elko fermions exhibit a range of intriguing mathematical and physical properties. To distinguish the essential features intrinsic to the theory from those that are merely artifacts of a particular basis choice, it is crucial to understand the underlying degrees of freedom. This line of investigation was initiated in Ref.~\cite{Ahluwalia:2022yvk,Ahluwalia:2023slc}, though the approach adopted there was not entirely satisfactory. 
Although the resulting fields are local and Lorentz covariant, it is unclear why the spinors $\xi_{\tau}(\p)$ and $\zeta_{\tau}(\p)$ are eigenspinors of an anti-linear charge-conjugation operator. We now understand that this is not the essential feature that determines the mass dimension and kinematics of Elko. 
The polarization spinors $u_{n}$ and $v_{n}$, which act as expansion coefficients for $\lambda(x)$, do not have to be eigenspinors of charge-conjugation. In this sense, the term Elko does not provide an accurate description of the theory. Nevertheless, we shall respect the historical development of the theory and continue to refer to the mass dimension one spinor field as Elko.
The crucial feature that distinguishes Elko (superposition formalism) from the Dirac field (doublet formalism) is the constraint on the relative phases in the polarizations, denoted by $b_{u,n}$ and $b_{v,n}$:
\begin{align*}
    \text{Elko: }& b_{u,+\frac{1}{2}}+b_{u,-\frac{1}{2}}=b_{v,+\frac{1}{2}}+b_{v,-\frac{1}{2}}=0\,. \\
    \text{Dirac: }&b_{u,+\frac{1}{2}} = b_{u,-\frac{1}{2}} 
= -b_{v,+\frac{1}{2}} = -b_{v,-\frac{1}{2}} = \pm 1\,.
\end{align*}
To be precise, as discussed at the end of Section~\ref{sec:wigner_dirac_field}, when the phases satisfy the Dirac condition, the Wigner-degenerate fermions cannot be described by the superposition field $\lambda(x)$ and must instead be described by the doublet field $\Psi(x)$. Once these phase relations are fixed, the appropriate kinematic structure follows straightforwardly.

Having established the foundation of the free theory, the next step is to study its interactions. Although the free Hamiltonian for Elko is Hermitian, due to the definition of the dual $\dual{\lambda}(x)$~\eqref{eq:elko_dual_0}, only a restricted class of interactions are Hermitian~\cite{deGracia:2024umr}. For example, the self-interaction $(\dual{\lambda}\lambda)^{2}$ and coupling to a $U(1)$ gauge field $J_{\mu}A^{\mu}$ where the current $J_{\mu}=ig[\dual{\lambda}\partial_{\mu}\lambda-(\partial_{\mu}\dual{\lambda})\lambda]$ are non-Hermitian. Specifically, these interactions are pseudo-Hermitian~\cite {Mostafazadeh:2001jk,Mostafazadeh:2008pw}. 
As the formalism for extracting physical observables from pseudo-Hermitian Hamiltonians is still under development, a comprehensive study of Elko interactions must await further theoretical advancements.
Meanwhile, several important problems are still open.
One notable issue concerns the massless limit of Elko.
For Elko to satisfy the canonical anti-commutation relations, the spinor fields must be rescaled by a factor of $\sqrt{m}$~\eqref{eq:elko_rescale_0}, rendering the theory singular as $m \to 0$. This suggests that Elko is intrinsically massive. 
From the perspective of modern particle physics, where masses arise from spontaneous symmetry breaking (SSB), a natural question arises: \textit{What is the origin of the Elko mass?} 
A promising direction is to explore chiral fermions with the Wigner degeneracy.
Such fermions satisfy the massless Dirac equation and are therefore described by the doublet field. The challenge then is to identify a viable SSB mechanism that can give rise to massive Elko fields.
In the conventional framework, chiral fermions acquire mass through Yukawa couplings to scalar fields.
Whether a similar mechanism can lead to the emergence of the Elko from a doublet field, potentially through a novel SSB process, remains an open and compelling question for future exploration.


\acknowledgments

CYL thanks Julio M. H. da Silva, Gabriel B. de Gracia and Rodolfo J. B. Rogerio for correspondence. SZ is supported by Natural Science Foundation of China under Grant No.12347101 and No.2024CDJXY022 at Chongqing University.

\appendix

\section*{Appendix}

\section{Notations \& conventions}
\label{app:notations_and_conventions}

Throughout the paper, we use the conventions of Ref.~\cite{Peskin:1995ev}.
The 4D Minkowski metric is,
\begin{equation}
\eta_{\mu \nu} = \eta^{\mu \nu} = \text{diag}(+1, -1, -1, -1) \, ,
\label{metric_1}
\end{equation}
where $\mu, \nu=0,1, 2, 3$.

The $4\times4$ Dirac matrices are taken in the Weyl representation,
\begin{equation}
\gamma^\mu =
\left[\begin{matrix}
0 & \sigma^\mu \\
\bar{\sigma}^\mu & 0
\end{matrix}\right] \, ,
\quad \text{with} \quad
\biggl\{
\begin{array}{r c l}
\sigma^\mu &=& \left( \mathds{1}_{2 \times 2}, \sigma^i \right) \, , \\
\bar{\sigma}^\mu &=& \left( \mathds{1}_{2 \times 2}, -\sigma^i \right) \, ,
\end{array}
\label{gamma_1}
\end{equation}
where $\mu=0,1,2,3$ and $\sigma^i$ ($i = 1, 2, 3$) are the three Pauli matrices:
\begin{equation}
\sigma^1 =
\left[\begin{matrix}
0 \quad  & \quad 1  \\
1 \quad  & \quad 0  
\end{matrix}\right] \, ,
\quad
\sigma^2 =
\left[\begin{matrix}
0 \quad  & -i \\
i \quad  & 0
\end{matrix}\right] \, ,
\quad
\sigma^3 =
\left[\begin{matrix}
1 \quad  &  0 \\
0 \quad  &  -1
\end{matrix}\right] \, ,
\end{equation}
and
\begin{align}
\left\{\sigma^{i}, \sigma^{j} \right\} = 2\delta^{ij} \, , \quad i,j = 1,2,3 \, .
\end{align}
One has also the chiral operator,
\begin{equation}
\gamma^5 = i \gamma^0 \gamma^1 \gamma^2 \gamma^3 =
\left[\begin{matrix}
- \mathds{1}_{2 \times 2} & 0 \\
0 & \mathds{1}_{2 \times 2}
\end{matrix}\right] \, ,
\label{gamma_2}
\end{equation}
and the chiral projection operators
\begin{equation}
P_{L,R} \equiv \dfrac{\mathds{1} \mp \gamma^{5}}{2} \, .
\end{equation}

\bibliography{Bibliography}
\bibliographystyle{JHEP}

\end{document}